%% file: main.tex
\documentclass[11pt]{article}

\usepackage[final]{main}

\usepackage{times}
\usepackage{latexsym}

\usepackage[T1]{fontenc}

\usepackage[utf8]{inputenc}

\usepackage{microtype}

\usepackage{inconsolata}

\usepackage{graphicx}

\usepackage{url}
\usepackage{booktabs} 
\usepackage{graphicx} 
\usepackage[table]{xcolor} 
\usepackage{colortbl}   
\usepackage{booktabs, multirow, colortbl, xcolor}
\usepackage{threeparttable}
\usepackage{lineno}
\usepackage{subcaption}
\usepackage{multicol}
\usepackage{wrapfig}
\usepackage{amsmath}
\usepackage{arydshln}
\usepackage[table]{xcolor}
\usepackage{multirow}
\usepackage{enumitem}

\usepackage[utf8]{inputenc} 
\usepackage[T1]{fontenc}    
\usepackage{hyperref}       
\usepackage{url}            
\usepackage{booktabs}       
\usepackage{amsfonts}       
\usepackage{nicefrac}       
\usepackage{microtype}      
\usepackage{xcolor}         

\usepackage{makecell}
\usepackage{adjustbox}

\definecolor{citecolor}{HTML}{0071bc}
\definecolor{lightgray}{gray}{0.9}
\definecolor{lightgreen}{rgb}{0.9,1,0.9}
\usepackage{pifont}

\newcommand{\cmark}{\textcolor{green!50!black}{\ding{51}}}%
\newcommand{\xmark}{\textcolor{red!50!black}{\ding{55}}}%
\newcommand{\authorskip}{\hspace{0.5mm}}

\usepackage{titlesec}
\titlespacing*{\section}{0pt}{6pt}{4pt}
\titlespacing*{\subsection}{0pt}{4pt}{2pt}
\titlespacing*{\subsubsection}{0pt}{4pt}{2pt}
\titlespacing*{\paragraph}{0pt}{4pt}{1em}

\title{\framework: An Open-Source Toolkit \\ for Efficient and Unified Evaluation of \underline{AU}dio-LLMs}

\author{%
\authorskip Hoang Nguyen$^{\S}$,
 Sidharth Surapaneni$^{\dag}$\thanks{Work done during internship at ServiceNow \\}, 
\textbf{Akshay Kalkunte$^{\S}$,}
 \textbf{Jash Mehta$^{\S}$,} 
 \textbf{Aman Tiwari$^{\S}$,} 
 \\ \textbf{Oluwanifemi Bamgbose$^{\S}$,} 
 \textbf{Khyati Mahajan$^{\S}$,}  \textbf{Jash Shah$^{\S}$,} \textbf{Shruthan Radhakrishna$^{\S}$,} \\ \authorskip \textbf{ Sathwik Tejaswi Madhusudhan$^{\S}$,} \textbf{Vikas Yadav$^{\S}$,} \textbf{Sai Rajeswar$^{\S}$}  \\
$^{\S}$ ServiceNow \\
$^{\dag}$ University of Texas at Austin \\
\small{
   \textbf{Correspondence:} \href{mailto:huuhoang.nguyen@servicenow.com}{huuhoang.nguyen@servicenow.com}
 }
}

\def\framework{AU-Harness}
\def \airbench{AIR-Bench}
\def \audiobench{AudioBench}
\def \voicebench{VoiceBench}
\def \kimieval{Kimi-Eval}

\def \metricrtf{RTF}

\def \vllm{vLLM}

\definecolor{ao}{rgb}{0.0, 0.5, 0.0}

\begin{document}
\maketitle
\begin{abstract}
Large Audio Language Models (LALMs) are rapidly advancing, but evaluating them remains challenging due to inefficient and non-standardized toolkits that limit fair comparison and systematic assessment. Existing evaluation frameworks exhibit three critical limitations: (1) slow and inefficient processing pipeline that bottlenecks large-scale studies, (2) inadequate multi-turn dialogue support, leaving fundamental questions about cross-turn context integration and performance dynamics over extended conversations in LALMs unanswered; and (3) the absence of unified and scalable evaluation framework capable of keeping pace with the rapid growth of both LALMs and audio benchmarks. To address these issues, we introduce \textbf{\framework}, an efficient and comprehensive evaluation framework for LALMs. 
Our empirical studies demonstrate that \framework~achieves 77\% reduction in evaluation processing time over existing evaluation toolkits, enabling large-scale LALM evaluations previously considered impractical.
We provide standardized evaluation protocols and flexible configurations for fair model comparison across diverse scenarios. \framework~unlocks a range of in-depth analyses difficult to conduct without a unified foundation, including multi-turn dialogue dynamics, enabling the study of true audio reasoning capabilities in existing LALMs.
\framework~provides both practical evaluation tools and insights into model limitations, advancing systematic LALM development.
\footnote{\href{https://github.com/ServiceNow/AU-Harness}{https://github.com/ServiceNow/AU-Harness}}
\end{abstract}

\input{sections/001_introduction}

\input{sections/002_related_work}
\input{sections/003_challenge}

\input{sections/004_framework}

\input{sections/005_discussion}
\input{sections/006_conclusion}


\bibliography{main}

\appendix
\input{sections/007_appendix}

\end{document}

%% file: sections/001_introduction.tex
\section{Introduction}
The emergence of Large Audio Language Models (LALMs) has opened new frontiers, extending capabilities beyond textual inputs to speech, sounds, and multimodal inputs~\citep{tangsalmonn,cui2024recent}. This progress has accelerated the development of frontier LALMs and audio-focused benchmarks. Recent multimodal LALMs like Gemini 2.5~\citep{comanici2025gemini}, Qwen2.5-Omni~\citep{xu2025qwen2} have demonstrated substantial audio understanding capabilities well beyond traditional Automatic Speech Recognition (ASR) tasks. However, despite these advances, audio evaluation toolkits~\footnote{We utilize the term evaluation \textbf{toolkit}, \textbf{kit}, and \textbf{framework} interchangeably across the manuscript.} have comparatively received little attention. Thus, there is a need for efficient, customizable, and consistent evaluation framework for fair model comparisons which can evolve as audio tasks and model complexities grow.

\input{tables/03_toolkit_comparison}

Existing efforts including \audiobench~\citep{wang-etal-2025-audiobench}, \kimieval~\citep{ding2025kimi}, VoiceBench~\citep{chen2024voicebench} and LMMS-Eval~\citep{zhang2025lmms} have provided extensive task coverage from ASR to spoken question answering and scene understanding.
However, prevailing toolkits still face three persistent limitations. First, \textbf{throughput}: many pipelines under-utilize batching and parallelism, creating bottlenecks that preclude large-scale, systematic comparisons. Second, \textbf{reproducibility}: ad-hoc prompting and non-standardized evaluation settings lead to incomparable performance across setups. Third, \textbf{task scope}: evaluations remain largely restricted to static single-turn interactions, failing to account for LALM assessment over extended interactions in multi-turn conversational settings which are more frequent in real world conversations.

Most current evaluation frameworks depend on simplistic yet inefficient input processing pipelines that struggle to scale with the increasing volume and complexity of audio benchmarks and LALMs. These limitations not only constrain the throughput of large-scale evaluations, but also hinder fair and reproducible comparisons across models of different sizes and architectures. As the field progresses toward more diverse and challenging audio tasks, the shortcomings of current evaluation infrastructure may pose a critical bottleneck, ultimately hampering the potential progress of LALMs. Unlike previous frameworks, we introduce an efficient token request orchestration together with effective data sharding to scale evaluations across multiple nodes and hardware architectures, leading to improved efficiency for audio benchmark evaluations. 

Beyond computational efficiency, existing toolkits exhibit critical limitations that collectively hinder systematic and reproducible LALM evaluation, as summarized in Table~\ref{tab:toolkit_comparison}. Despite varying in scope and design, no existing framework provides comprehensive coverage across the key dimensions required for rigorous evaluation: efficient and comprehensive inference backend integration, multi-turn dialogue support, concurrent multi-task execution, and configurable evaluation design. Together, these limitations force researchers to either compromise evaluation rigor or resort to fragmented, toolkit-specific workarounds that undermine reproducibility and cross-benchmark comparability. \framework~is the first unified and efficient toolkit to provide comprehensive support across all five dimensions, enabling systematic and reproducible LALM evaluation at scale. Our contributions are summarized as follows:
\vspace{-\topsep}
\begin{itemize} [nolistsep, noitemsep, topsep=0pt, partopsep=0pt, leftmargin=*]
    \item We propose an \textbf{efficient evaluation engine} that leverages \vllm~batching and dataset sharding to scale evaluations to multi-node infrastructures without sacrificing fidelity.
    
    \item A \textbf{unified, configurable framework} that standardizes evaluation protocols across benchmarks, enabling fair, reproducible comparisons and easy task integration.

    \item To the best of our knowledge, \textbf{\framework} is the first unified evaluation framework to make \textbf{multi-turn} dialogue evaluation fully configurable, enabling systematic analyses of dialogue behaviors over extended interactions.    
\end{itemize}

%% file: tables/03_toolkit_comparison.tex
\begin{table*}[tbh]
\centering
\resizebox{\textwidth}{!}{%
\begin{tabular}{lccccccc@{}}
\toprule
\textbf{EvalKit} & \textbf{HF Support} & \textbf{vLLM Support} & \textbf{Multi-task Parallel} & \textbf{Multi-turn} & \textbf{Customizable} & \textbf{RTF $\downarrow$} & \textbf{SPS $\uparrow$} \\
\midrule
\textbf{\audiobench}~\cite{wang-etal-2025-audiobench}  & \cmark & \xmark & \xmark & \xmark & \xmark & 101.29 $\pm$ 6.37 & 0.34 $\pm$ 0.02 \\
\textbf{\kimieval}~\cite{ding2025kimi}    & \cmark & \xmark & \xmark & \xmark & \xmark & 54.13 $\pm$ 0.07  & 0.63 $\pm$ 0.01 \\
\textbf{\voicebench}~\cite{chen2024voicebench}  & \cmark & \xmark & \xmark & \xmark & \xmark & 129.67 $\pm$ 5.30 & 0.29 $\pm$ 0.03 \\
\textbf{LMMs-Eval}~\cite{zhang2025lmms}    & \cmark & \cmark & \cmark & \xmark & \xmark & 20.84 $\pm$ 0.62  & 1.63 $\pm$ 0.05 \\
\textbf{AHELM}~\cite{lee2025ahelm}        & \cmark & \cmark & \xmark & \cmark & \xmark & 57.86 $\pm$ 14.69  & 0.62 $\pm$ 0.18 \\
\midrule
\textbf{\framework}   & \cmark & \cmark & \cmark & \cmark & \cmark & \textbf{10.29 $\pm$ 0.61} & \textbf{3.32 $\pm$ 0.15} \\
\bottomrule
\end{tabular}%
}
\captionsetup{font=small}
\caption{\textbf{Feature comparison of contemporary LALM evaluation toolkits.} 
We evaluate key technical capabilities across existing frameworks: vLLM 
integration for efficient batching, HuggingFace (HF) model support, multi-turn 
dialogue support for conversational scenarios, parallel processing support for 
multi-task concurrent evaluation, and configurable customizations for flexible 
evaluation design. Throughput is further reported on MELD-Emotion~\cite{poria2019meld} via Samples Per Second 
(SPS$\uparrow$) and Real-Time Factor (RTF$\downarrow$) averaged across 3 runs to empirically validate 
efficiency claims beyond feature support alone. Refer to \S\ref{subsec:challenge_efficiency} for further details. \textbf{Our framework provides comprehensive support across most dimensions.}}
\label{tab:toolkit_comparison}
\end{table*}

%% file: sections/002_related_work.tex
\section{Related Work}
\paragraph{Audio Benchmarks.}
Benchmarks play a critical role in the development of LALMs. SUPERB~\citep{yang2021superb} established core task axes (Content, Speaker, Semantics, Paralinguistics) for audio model evaluation. DynamicSUPERB~\citep{huang2024dynamic} and DynamicSUPERB-2.0~\citep{huangdynamic} expanded coverage to instruction-tuned and sequence generation tasks across speech, music, and environmental audio. Instruction-following and agentic conversational behaviors have been further probed by \airbench~\citep{yang2024air} and VoiceBench~\citep{chen2024voicebench}. More recently, \audiobench~\citep{wang-etal-2025-audiobench} unifies 8 task families over 26 datasets for LALMs.

Recent complementary efforts broaden the breadth and depth with audio reasoning capabilities: X-ARES~\citep{xares2025} systematically assesses general audio encoders across domains, MECAT~\citep{mecat2025} targets fine-grained audio understanding with expert-guided captions and QA; MMAR~\citep{ma2025mmar}, MMAU-PRO~\citep{kumar2025mmau}, and MMSU~\citep{wang2025mmsu} focus on understanding and analyzing complex audio scenes, spatial relationships, and mixed-audio reasoning. 
Yet despite the rapid growth of audio benchmarks, the development of audio evaluation frameworks for fair and consistent comparisons between frontier models and benchmarks remains fairly understudied. This critical gap necessitates the development of a unified and efficient evaluation engine designed specifically for scalable audio evaluations.

\paragraph{Audio Evaluation Kits.} In contrast with Audio Benchmark development, Audio Evaluation Kits have received less attention. This can be primarily attributed to the straightforward nature and minimal setup requirements of the early audio tasks, as presented in Dynamic-SUPERB-2.0~\citep{huangdynamic} and AIR-Bench~\citep{yang2024air}. However, the rapid growth of LALMs and the increasing complexity of newly curated audio benchmarks have underscored the critical need for comprehensive evaluation kits, as exemplified through the recent development of extensive evaluation kits ~\citep{ding2025kimi,wang-etal-2025-audiobench,zhang2025lmms}. For instance, \audiobench~\citep{wang-etal-2025-audiobench} offers versatile evaluation support for up to 8 tasks across 26 datasets. VERSA~\citep{shi2025versa} introduces a comprehensive framework to evaluate the quality of various speech, audio and music signals, with the focus on text-to-audio applications. AHELM~\citep{lee2025ahelm} adds support for multi-turn audio conversation evaluation; however, its benchmark suite is largely fixed, and extending it to unsupported tasks requires writing task-specific code rather than specifying evaluations through a customizable configuration interface.
Despite these advancements, most current evaluation kits operate on the simplified assumption that \textit{a single model is evaluated against a single benchmark per run}. Addressing this limitation, we introduce an efficient, customizable evaluation kit to support the massive scale of the current LALMs and audio benchmarks as summarized in Table \ref{tab:toolkit_comparison}.

%% file: sections/003_challenge.tex
\section{LALM Evaluation Challenges}
\label{sec:challenge}

\subsection{Inference Efficiency}
\label{subsec:challenge_efficiency} 
Most existing LALM evaluation kits have been designed based on the assumption that \textit{a single model should be evaluated against a single benchmark per run}. However, this constrains researchers from conducting systematic, large-scale comparisons across LALMs and audio benchmarks efficiently, slowing the iterative process of model development and refinement. Current evaluation kits also under-utilize parallel processing capabilities available in the high-performance computing clusters, resulting in failures in incorporating benefits of available hardware infrastructures.

Two essential task-agnostic metrics for assessing the efficiency of LALM evaluation frameworks are \textit{Real-time Factor (RTF)} and \textit{Samples Processed per Second (SPS)}. RTF measures the processing time of an evaluation framework relative to the duration of the processed audio \cite{arriaga2024evaluation}. Lower RTF is more desirable, indicating a more efficient audio evaluation framework. On the other hand, SPS directly quantifies the model's processing speed by measuring the average number of audio samples processed per second. It serves as a complementary measure to RTF, providing a more granular view of the model's throughput and computational efficiency. The detailed formulation of these metrics are provided in Appendix \ref{app:rtf_details}.

To quantify the efficiency of existing evaluation frameworks, we conduct a study on $N=500$ audio samples (approximately 0.41 hours) of MELD-Emotion dataset~\citep{poria2019meld} with experimental details in Appendix \ref{appendix:efficiency_setup}. As reported in Table \ref{tab:toolkit_comparison}, most existing kits exhibit high RTF and low SPS, revealing fundamental throughput bottlenecks that ultimately hinder rapid LALM assessment over more diverse benchmark suites.

\subsection{Customizable Evaluation Configurations}
\paragraph{Multi-turn Dialogue Support} 
Previous audio evaluation toolkits have largely been constrained to tasks centered on single-turn user interactions. However, as the field moves toward building interactive and context-aware voice assistants, the ability to evaluate multi-turn tasks becomes increasingly critical. Multi-turn evaluation enables a more realistic assessment of dialogue continuity, contextual reasoning, and the model’s capacity to adapt dynamically across extended conversations. Without such support, current evaluation approaches risk overlooking key aspects of usability and robustness that are essential for next-generation LALMs in realistic agentic voice systems.

\paragraph{Evaluation Customization.} The lack of customizable filtering poses a significant barrier for researchers aiming to conduct in-depth analyses of current LALM limitations. Without the ability to refine evaluation datasets based on specific criteria, it is challenging to gain granular understanding of model performance across diverse audio conditions. For instance, while certain LALMs might perform reliably on 10-second audio chunks, they might be unable to handle short-form audio typically encountered in dialogue-state tracking systems.

%% file: sections/004_framework.tex
\input{layout_figures/overview_framework_fig}
\section{\framework}
\label{sec:framework}
In response to the presented challenges, we propose a standardized, efficient, highly customizable evaluation framework, \textbf{\framework}, detailed in Figure \ref{fig:overview_lalmeval}. \framework~is composed of 3 primary components: \textbf{Config}, \textbf{Central Request Controller (CRC)} and \textbf{Concurrent Engines}. The Config module defines a structured and hierarchical representation of customizable configurations, enabling flexible and transparent evaluation settings. The CRC is responsible for managing token requests and coordinating execution across the framework. Finally, the Concurrent Engines module carries out task-specific evaluations in parallel, where each engine can support multi-model evaluations tailored to particular tasks. In the following sections, we introduce our architecture design in detail to address the presented challenges in Section \ref{sec:challenge}.

\subsection{Inference Efficiency}
\label{section_4_1:inference_efficiency}

As illustrated in Figure~\ref{fig:overview_lalmeval}, \framework~maximizes inference throughput through three complementary mechanisms. First, we introduce  a \textit{Central Request Controller} (CRC) --- a token-based scheduling architecture that maintains a global pool of concurrency slots shared across all models and evaluation engines. Each slot represents permission to issue one inference request; slots are acquired before dispatch and released upon completion, ensuring that throughput is governed solely by user-defined global request limits rather than model or engine-specific constraints. User-specified retry counts further provide a tunable balance between throughput and reliability, automatically re-attempting failed requests without manual intervention. Second, \framework~employs a \textit{layered request synchronization} strategy that adaptively staggers wait times across concurrent models, increasing the probability that models processing the same dataset segment complete inference in a temporally aligned manner; minimizing idle periods and intra-engine waiting time. Third, \textit{proportional dataset sharding} partitions evaluation data into disjoint subsets distributed across model endpoints in proportion to each endpoint's concurrency capacity, enabling improved throughput scaling across heterogeneous resources (as studied in Appendix \ref{appendix:inference_efficiency_ablations} and Figure \ref{fig:3_efficiency_experiments}). Together, these mechanisms deliver scalable, predictable evaluation throughput with minimal engineering overhead.

\subsection{Customizable Evaluation Configurations}
\paragraph{Backend-Agnostic Inference}
\framework~ decouples predictive inference and metric computation from model hosting, requiring only a standardized model specification to integrate with the 
evaluation pipeline, regardless of whether the model is served via \vllm, a third-party API, or a custom FastAPI~\citep{fastapi} endpoint. This design provides native support for \vllm-compatible models for high-throughput inference, while remaining fully compatible with any backend exposing a standard \texttt{/v1/chat/completions} interface. To lower the potential integration barriers, \framework~provides boilerplate FastAPI server implementations, allowing practitioners to wrap any optimized inference stack with minimal overhead and seamlessly integrate into the evaluation pipeline.

\begin{figure*}[htb]
    \centering
    \includegraphics[width=0.95\linewidth]{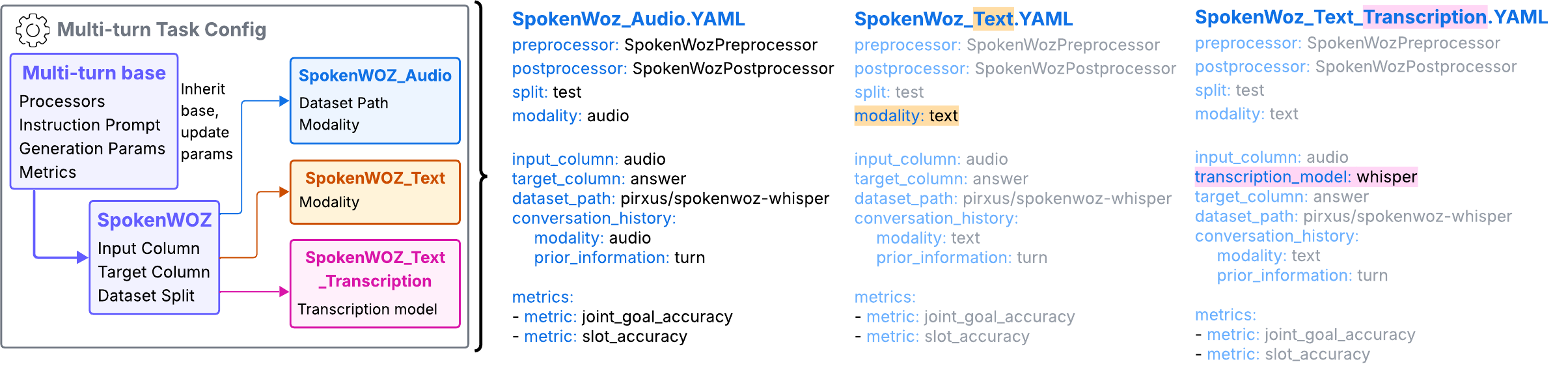}
    \caption{\small{\textbf{Compositional task configuration in \framework, illustrated on 
SpokenWOZ.} A shared base config defines all invariant components --- 
processors, instruction prompt, generation parameters, and metrics --- 
inherited by all task-specific variants. Each variant requires modifying 
only a single targeted field, such as input modality or transcription 
strategy, enabling new evaluation scenarios to 
be instantiated in minutes without any pipeline re-engineering. This 
compositional design is representative of \framework's broader 
configuration system, where comprehensive coverage of diverse tasks and 
settings is achieved through minimal, targeted modifications; 
facilitating rapid, reproducible experimentation at scale.}}
    \label{fig:task-config}
\end{figure*}

\paragraph{Evaluation customization.}
\framework~is also designed for granular control over evaluation steps. First, \framework~supports both open-source and proprietary models, which might contain their individualistic settings. Second, we allow for customizable metric assignment on a per-dataset and/or per-task basis. For instance, LLM-as-judge supports configurable concurrency to maximize the throughput for evaluation stage. For a more comprehensive understanding of model performance, the framework offers configurable aggregation metrics. This capability allows for the multi-dimensional analysis of task and metric results, providing a comprehensive outlook that extends beyond simple, individual scores or sub-tasks. 

As shown in Figure \ref{fig:overview_lalmeval}, evaluation behavior is  entirely specified through a YAML configuration, including the task–wise metrics, model configurations, optional score aggregation across related tasks, and prompt overrides.
Importantly, no task-specific Python code or glue logic is required, even for complex benchmarks such as Emotion Recognition task suite that involves aggregation of multiple sub-datasets and LLM-based judging.
This abstraction cleanly decouples tasks, metrics, models, and judges, enabling new evaluations to be launched by editing a single configuration file. User-friendly configuration is further illustrated in Appendix \ref{sec:au-harness-config-ui}.

\subsection{Support for features and tasks}
\paragraph{Multi-turn Dialogue support.}
Unlike existing toolkits, \framework~supports multi-turn evaluation in both audio and text modalities across LALMs through synchronous, turn-based evaluation chains that recursively append model outputs to the dialogue context at each turn. Beyond simple turn chaining, multi-turn audio evaluation introduces a non-trivial design space that existing toolkits leave unaddressed: from the second turn onward, the dialogue history may contain mixed-modality content, and the representation of prior turns must be explicitly configured. Specifically, each historical turn can be represented as 
(1) the original audio recording, (2) a ground-truth or reference text transcript, or (3) the model's generated transcription from the previous pass. The optimal choice is not universal --- it depends on the modality support and architectural constraints of the evaluated model. \framework~addresses this by providing dynamic configuration of history representation, enabling controlled and reproducible multi-turn evaluations across heterogeneous LALMs without requiring custom pipeline modifications for each model or dataset combination. This flexible design is further demonstrated via Section \ref{sec:5_multiturn}.

\paragraph{Continual engagement with emerging tasks.} We aim to continue supporting up-and-coming benchmarks and models so \framework\ can continue providing insights into LALM performance as they evolve. Figure \ref{fig:task-config} shows the ease of adding new datasets and evaluation configurations given the task category - the setup for a new task requires minimal updates with the given base configurations to be supported in \framework.

%% file: layout_figures/overview_framework_fig.tex
\begin{figure}[ht]
    \captionsetup{font=small}
    \centering
   \includegraphics[width=\columnwidth]{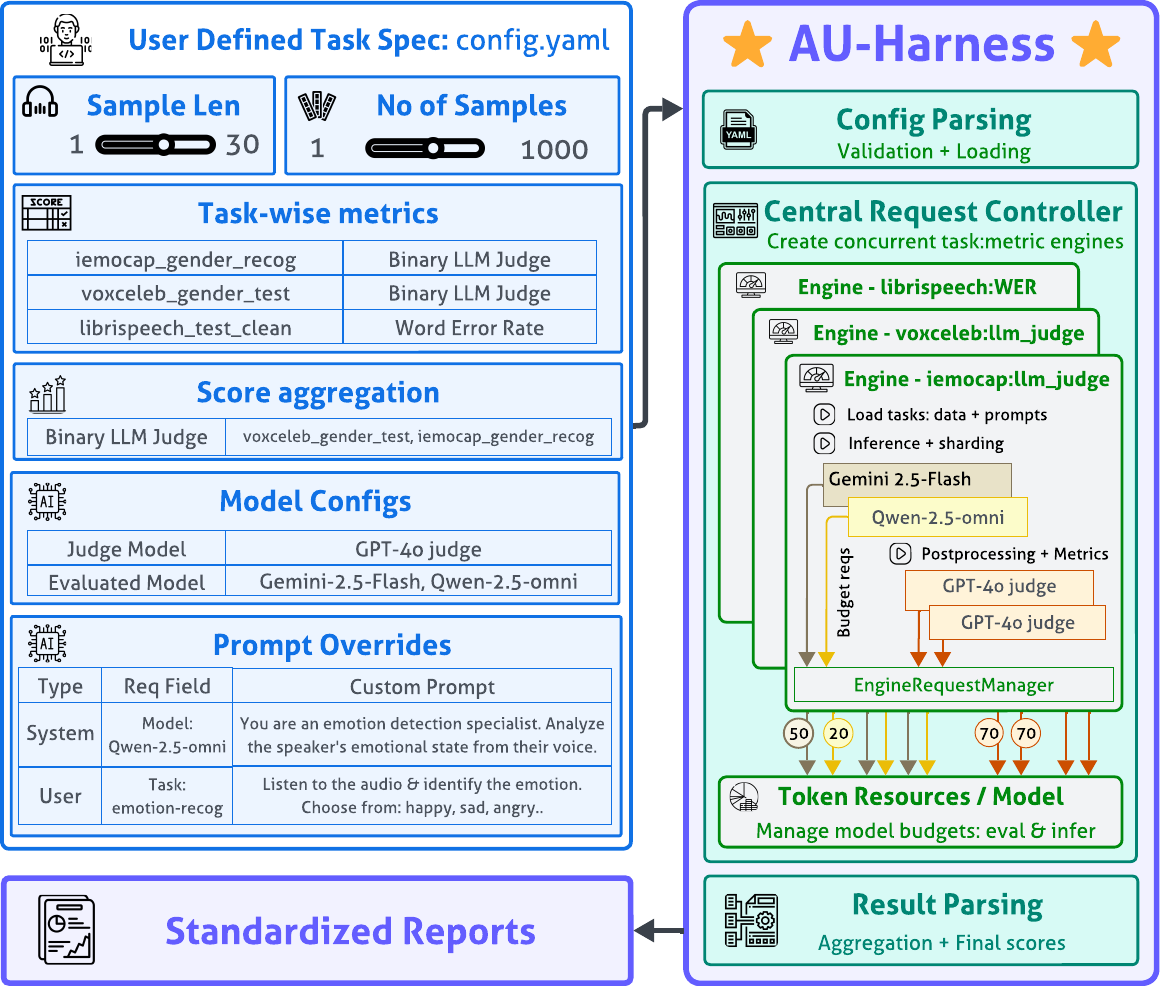}
    \caption{\textbf{Architecture overview of \framework{} evaluation framework.} Our system comprises three core components: (1) \textit{Config} module for hierarchical task configuration and standardized prompting, (2) \textit{Central Request Controller (CRC)} managing token-based concurrency limits across all engines with adaptive retry mechanisms, and (3) \textit{Concurrent Engines} executing parallel model evaluation with dataset sharding. The Central Request Controller maintains a global Token Pool accessible to all engines, enabling efficient resource utilization and scalable throughput. Multiple concurrent connections between the controller and inference models illustrate parallel request dispatch, with each engine supporting the evaluation of multiple models on targeted datasets.}.
    \label{fig:overview_lalmeval}
\end{figure} 

%% file: sections/005_discussion.tex
\input{tables/05_main_benchmark_v2}

\section{ Results \& Discussion}
\label{sec:results}
Without loss of generality, we adopt the task taxonomy proposed by Dynamic-SUPERB-2.0~\cite{huangdynamic} for the empirical evaluations with our proposed \framework~due to its exhaustive coverage.  Table \ref{tab:main_updated_taxonomy_benchmark} characterizes the breadth of audio evaluation suite supported by our \framework, demonstrating the flexibility of our \framework~in supporting diverse audio tasks.  Following \cite{wang-etal-2025-audiobench}, we adopt GPT-4o-mini as judge for LLM-judge metrics \footnote{Results are obtained prior to the API discontinuation date} due to its advanced capability. Further details of datasets, evaluated models and metrics are provided in Appendix \ref{appendix:comprehensive_eval}. 

\subsection{Inference Efficiency}

\label{subsec:inference_efficiency}
\paragraph{Evaluation Settings.} 
We benchmark \framework~against AudioBench~\citep{wang-etal-2025-audiobench}, 
VoiceBench~\citep{chen2024voicebench}, Kimi-Eval~\citep{ding2025kimi}, and 
LMMS-Eval~\citep{zhang2025lmms} on 500 audio samples across three diverse datasets 
and three  LALMs, reporting averaged \metricrtf~and SPS. Additional runtime setups, namely \textit{Sequential}, \textit{Total} and \textit{Parallel}~\footnote{Parallel refers to Parallel(Optimal) where no actual overhead is accounted for unless specified otherwise.}, to assure a comprehensive and fair comparison among all existing evaluation kits are also examined as detailed in Appendix \ref{appendix:efficiency_setup}.

\input{layout_figures/runtime_bar_plot}

\paragraph{Evaluation Comparison.}
As shown in Figure \ref{fig:runtime_barplot}, \framework~outperforms existing evaluation kits across different runtime scenarios in two key efficiency metrics. Against the most competitive baseline, LMMS-Eval with \vllm{} backend, \framework~achieves up to 77.2\% relative reduction in RTFs under realistic end-to-end conditions (\textit{Total} runtime). This improvement, as further demonstrated in Figure~\ref{fig:runtime_scatterplot}, reflects substantial throughput gains when compared to contemporary evaluation frameworks across all evaluated datasets and models.
The proposed orchestration layer (Figure \ref{fig:overview_lalmeval}) is designed to interface with any backend supporting dynamic scheduling, optimizing request handling and token delegation to minimize idle waiting time. As such, it is readily transferable to any inference backend sharing \vllm's key characteristics: continuous batching, asynchronous request processing, and dynamic memory management.

\paragraph{Orchestration pipeline beyond \vllm~integration.} To validate the efficiency gains of CRC beyond naive \vllm~integration, we conduct a comparative study against LMMS-Eval \cite{zhang2025lmms}, a framework that similarly leverages \vllm~for large-scale multimodal evaluation. The critical distinction lies in the scheduling strategy: while LMMS-Eval relies on \vllm's default processing pipeline with multi-threaded execution only, \framework~introduces dynamic concurrent request orchestration that actively minimizes idle waiting time across tasks. As shown in Table \ref{tab:ablation_efficiency_compare_lmms}, although LMMS-Eval's naive \vllm~integration already improves throughput over HF-based counterparts (42.5\% average SPS relative gain), it remains substantially less efficient than \framework, which achieves an additional approx. 72.3\% average SPS relative improvement. This gap widens further under realistic concurrent task processing conditions (\textit{Reality}), where \framework~outperforms LMMS-Eval by 51.2\% in RTF relative reduction and 109.2\% in average SPS relative gain — demonstrating that effective \vllm~utilization requires deliberate orchestration beyond default pipeline configurations.

\input{tables/05_ablation_compare_lmms}
\subsection{Insights enabled by \framework}
\label{sec:5_2_insight}
A unified evaluation framework does more than evaluating models efficiently and comprehensively as it enables systematic analyses that would be difficult or impossible to conduct with fragmented toolkits. \framework~presents these through an interactive run report (Appendix~\ref{sec:au-harness-run-report}, Figures~\ref{fig:rr-overview}--\ref{fig:rr-ops-health}) covering category performance, error patterns, multi-turn dynamics, head-to-head comparison, and operational health. This enables further analyses including instruction modality effects, and multi-turn dialogue dynamics. Rather than an exhaustive survey, these case studies are intended to illustrate the analytical depth that a unified, extensible framework provides, and to highlight open challenges in the rigorous LALM evaluations.

\input{tables/05_ablation_cascade_v2}
\paragraph{Instruction Modality Gap.} \label{sec:instruction-modality}
Table~\ref{tab:modality_cascaded} reveals two compounding effects that expose fundamental limitations in current LALM evaluation practices. First, the modality of the instruction itself introduces measurable degradation: even under the strongest ASR condition (Whisper-v3, BERTScore 99.59), IFEval performance drops from 87.56 to 79.74 ($\Delta$ = -7.82), confirming that a residual instruction modality gap persists independent of transcription quality. Second, and more critically, the choice of evaluation paradigm dramatically shapes what benchmarks measure. On IFEval, the model's own ASR achieves only 37.52 BERTScore and collapses task performance to 30.07 (
$\Delta$ = -57.49); yet even Qwen3-Captioner at 88.57 BERTScore only partially recovers performance to 58.76 (
$\Delta$ = -28.80), demonstrating that transcription quality alone does not predict task recovery. The pattern reverses on GSM8K: direct audio reasoning severely degrades performance (
$\Delta$ = -30.93), while Own ASR preprocessing recovers it to 72.02 (
$\Delta$ = -5.31) and Whisper-v3 fully closes the gap (
$\Delta$ = $+0.46$). This task-dependent asymmetry reveals that the cascaded intelligence bottleneck is not uniform: while instruction-following tasks are disproportionately sensitive to transcription artifacts, mathematical reasoning is primarily bottlenecked at the audio-grounded inference stage rather than the transcription stage. Uncovering these distinctions is feasible through a unified evaluation framework that supports
modular pipelines and standardized evaluation protocols across diverse modality and inference conditions as offered in \framework. 

\paragraph{Multi-turn Dialogue Dynamics.}
\label{sec:5_multiturn}

\begin{figure*}[htb]
    \centering
    \captionsetup{font=small}
   \includegraphics[width=0.8\linewidth, trim=0.0cm 0.0cm 0cm 42pt, clip]{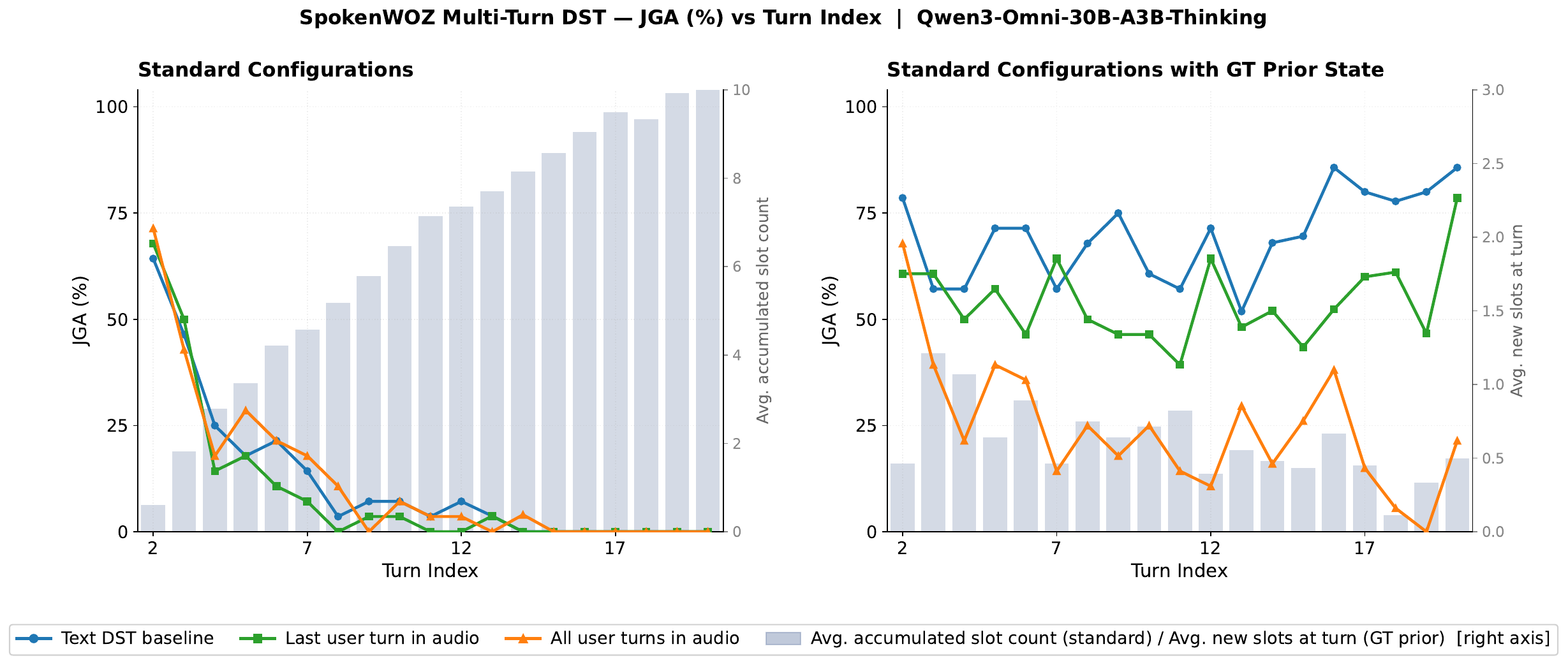}
    \caption{\textbf{SpokenWOZ multi-turn DST --- JGA vs.\ turn index across input-format configurations for Qwen3-Omni-30B-A3B-Thinking.} Each panel groups evaluation configurations enabled by \framework\ by  scoring strategy: \textit{Standard Configurations} (left) vary the current-turn modality across text-only (baseline), last user turn in audio, and all user turns in audio, evaluated against the cumulative dialogue state; \textit{Standard Configurations with GT Prior State} (right) apply the same modality variants but inject the oracle prior state before each turn, scoring only the per-turn delta. Grey bars denote the mean accumulated slot count per turn (left panel) and the mean number of new slots introduced per turn (right panel), both on the right axis. Further details in Section \ref{sec:turn-range-selection}.}
    \label{fig:spokenwoz_multiturn}
\end{figure*} 

Figure~\ref{fig:spokenwoz_multiturn} presents JGA trajectories across six input-format configurations on SpokenWOZ, each enabled by \framework's flexible multi-turn evaluation design. Across all standard configurations, JGA declines sharply with each dialogue turn and approaches zero by turn~14. This trend closely tracks the monotonically increasing number of accumulated slots to predict, indicating that cumulative slot complexity is a major driver of performance collapse. The oracle configurations isolate this effect by providing the ground-truth prior state and scoring only per-turn slot deltas, thereby removing accumulated slot errors as a confounder. Under this setting, JGA stabilizes substantially, suggesting that unfilled slot accumulation, not turn-level understanding, is the dominant bottleneck. The oracle curves therefore provide a more faithful estimate of per-turn comprehension capacity. Text-only inputs achieve the highest JGA in both standard and oracle settings, consistent with strong transfer from extensive text-based pretraining in LALMs. Presenting the current turn in audio lowers JGA relative to the text-only baseline, revealing a persistent audio-text perception gap even when only the current turn is spoken. This gap becomes more pronounced under the oracle setting, suggesting that it is partially masked by error accumulation in standard cumulative-state evaluation. The all-audio configuration, in which all user turns are spoken, exhibits an ambiguous trend under standard evaluation but shows substantially lower JGA under the oracle setting, exposing a more fundamental audio-only comprehension gap. Crucially, this analysis spans modality, history representation, and state-tracking strategy within a single evaluation framework. \framework's composable, per-model multi-turn configuration system enables these comparisons directly, whereas existing toolkits would require bespoke pipeline engineering.

%% file: tables/05_main_benchmark_v2.tex
\begin{table*}[htbp!]
\centering
\resizebox{\textwidth}{!}{%
\begin{tabular}{@{}lccccccccccccc@{}}
\toprule
\multicolumn{1}{c}{\textbf{Models}} & \multicolumn{11}{c}{\textbf{Speech}} & \multicolumn{2}{c}{\textbf{Audio \& Music}} \\ \cmidrule(lr){1-1} \cmidrule(lr){2-12} \cmidrule(lr){13-14}
\textbf{Task Category} & \textbf{PR} & \textbf{ASR} & \textbf{Para.} & \textbf{S\&L} & \textbf{SLU} & \textbf{HD} & \textbf{SE} & \textbf{Safety} & \multicolumn{2}{c}{\textbf{Multi-turn}} & \textbf{IF} & \textbf{AU} & \textbf{MU} \\
Dataset & voxangeles & Librispeech & IEMOCAP & SR & BBA & StutterDetect & NoiseDetect & Advbench & MT-Bench & SpokenWoz & IFEval & AudioCaps & ChoMusic \\
Metrics & LB (↑) & WER (↓) & LB (↑) & LB (↑) & LBBA(↑) & LB (↑) & LB (↑) & SafetyJudge (↑) & MTJudge (↑) & JGA (↑) & IFScore (↑) & LB (↑) & LB (↑) \\
\midrule

\rowcolor[HTML]{A2EBCD}
\multicolumn{14}{c}{\textbf{Small-sized Audio Language Models (\textless{}5B parameters)}} \\
\rowcolor[HTML]{F0FAF6}
Voxtral-Mini-3B & 0.10 & 2.10 & 54.90 & 45.80 & 43.50 & 12.90 & 14.50 & 78.50 & 65.88 & 24.92 & 40.02 & 14.96 & 45.40 \\
\rowcolor[HTML]{F0FAF6}
Qwen2.5-Omni-3B & 1.20 & 8.09 & 81.50 & 55.90 & 44.80 & 58.40 & 58.40 & 97.30 & 59.81 & 31.30 & 35.82 & \textbf{42.82} & 53.50 \\

\rowcolor[HTML]{A8CDF7}
\multicolumn{14}{c}{\textbf{Medium Sized Large Audio Language Models (5B--20B parameters)}} \\
\rowcolor[HTML]{F0F6FD}
Phi-4-Multi-modal & 0.00 & 1.97 & 50.50 & 47.20 & 40.80 & 42.10 & 32.50 & 97.10 & 64.12 & 7.82 & 44.51 & 26.08 & 44.80 \\
\rowcolor[HTML]{F0F6FD}
Qwen2.5-Omni-7B & 21.60 & 1.74 & 85.80 & 62.30 & 50.90 & \underline{68.20} & 59.00 & 98.30 & 64.56 & \underline{37.30} & 50.83 & 38.40 & 59.30 \\
\rowcolor[HTML]{F0F6FD}
Kimi-Audio & 1.30 & \textbf{1.41} & \underline{89.00} & \underline{62.80} & 41.70 & 58.40 & \textbf{96.00} & \textbf{100.00} & 54.62 & \underline{37.30} & 61.29 & \underline{38.46} & 66.80 \\

\rowcolor[HTML]{CFC9FF}
\multicolumn{14}{c}{\textbf{Large Sized Large Audio Language Models (\textgreater{}20B parameters)}} \\
\rowcolor[HTML]{F5F4FE}
Voxtral-Small-24B & 1.20 & \underline{1.62} & 42.80 & 47.70 & 66.50 & 51.90 & 15.50 & 75.40 & 70.81 & 29.93 & 66.83 & 19.24 & 57.90 \\
\rowcolor[HTML]{F5F4FE}
\shortstack[l]{Qwen3-Omni-30B \\ -A3B-Thinking} & \textbf{50.90} & 1.64 & 82.50 & \textbf{65.30} & \textbf{96.80} & \textbf{68.70} & 78.00 & 95.00 & \textbf{76.19} & 11.16 & \underline{80.39} & 37.96 & \textbf{74.30} \\

\rowcolor[HTML]{F7C1B0}
\multicolumn{14}{c}{\textbf{Proprietary Audio Language Models}} \\
\rowcolor[HTML]{FEF6F3}
GPT-4o-mini-audio $^\dagger$ & 0.00 & 6.25 & --* & 40.30 & 63.70 & 3.80 & 53.00 & 88.10 & 65.00 & 28.39 & 70.47 & 15.08 & 50.20 \\
\rowcolor[HTML]{FEF6F3}
Gemini2.5-Flash & 35.80 & 2.17 & \textbf{92.70} & 60.20 & \underline{90.30} & 64.60 & \underline{87.50} & \underline{98.50} & \underline{74.50} & \textbf{52.09} & \textbf{84.63} & 36.16 & \underline{72.90} \\

\rowcolor[HTML]{B4B2A9}
\multicolumn{14}{c}{\textbf{Cascaded Systems}} \\
\rowcolor[HTML]{F7F6F4}
\shortstack[l]{Whisper-Large-v3 + \\ GPT-oSS-20B} & \underline{48.00} & 9.82 & 1.90 & 48.10 & 78.20 & 49.80 & 49.00 & \underline{98.50} & 71.50 & 19.02 & 73.72 & 12.14 & 50.30 \\
\rowcolor[HTML]{F7F6F4}
\shortstack[l]{GPT-4o-transcribe + \\ GPT-4.1-mini}$^\dagger$ & 30.70 & 4.71 & 30.20 & 45.80 & 74.30 & 49.70 & 47.00 & 97.30 & 67.62 & 19.44 & 66.69 & 17.44 & 52.70 \\

\bottomrule
\end{tabular}%
}
\captionsetup{font=small}
\caption{\textbf{LALM performance on audio tasks.} We evaluate representative LALMs from different spectra: Open-source LALMs(small-sized, medium-sized, large-sized), Proprietary LALMs and Cascaded System LALMs across representative task categories. Metrics include LLM-as-judge evaluations using GPT-4o-mini and task-specific automatic metrics. \textbf{PR}: Phoneme Recognition, \textbf{Para.}: Paralinguistics, \textbf{ASR}: Automatic Speech Recognition, \textbf{S\&L}: Speaker \& Language, \textbf{SLU}: Spoken Language Understanding, \textbf{HD}: Hearing Disorder, \textbf{SE}: Speech Enhancement, \textbf{IF}: Instruction Following, \textbf{AU}: Audio Understanding, \textbf{MU}: Music Understanding, Refer to Appendix \ref{appendix:comprehensive_eval} for benchmark, model and metric abbreviations together with their detailed explanations. \textbf{Bold}: highest; \underline{underline}: second highest. *Performance affected by Azure OpenAI content filtering. $^\dagger$ denotes results obtained prior to the API discontinuation date}
\label{tab:main_updated_taxonomy_benchmark}
\end{table*}

%% file: layout_figures/runtime_bar_plot.tex
\begin{figure*}[htb]
    \centering
    \captionsetup{font=small}
    \subfloat[Real-time Factor (RTF, $\downarrow$ better)]
    {{\includegraphics[trim={0 0 0 0.6cm}, clip, width=.40\textwidth]{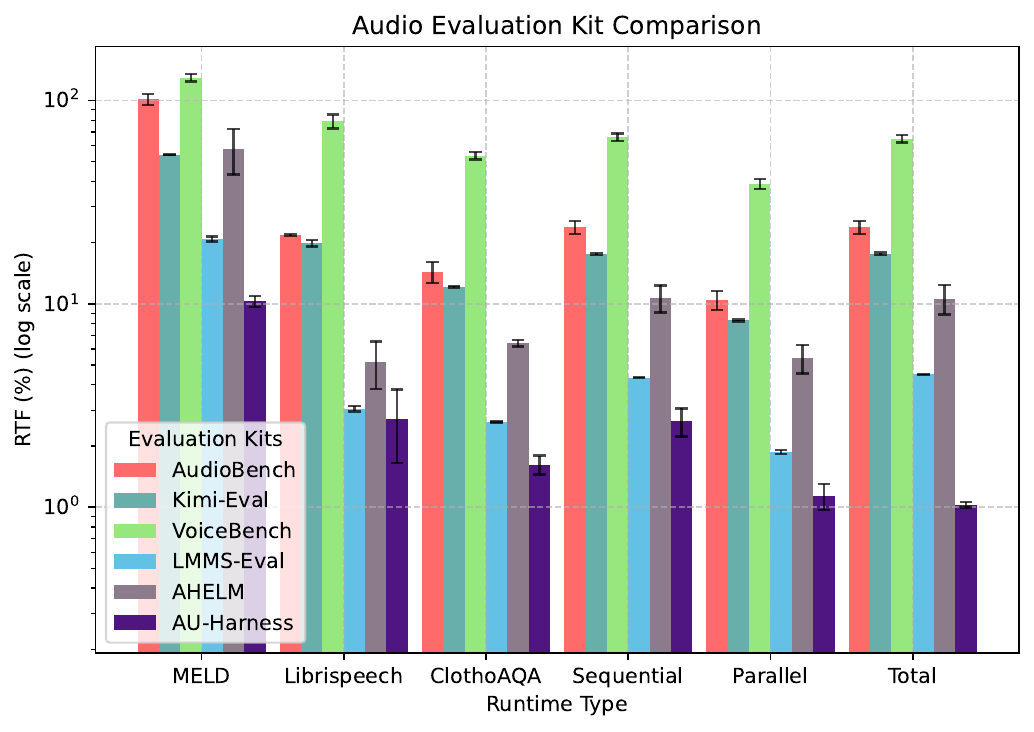} }}%
    \qquad
    \subfloat[Samples Processed per Second (SPS, $\uparrow$ better)]
 {{\includegraphics[trim={0 0 0 0.6cm}, clip,width=.40\textwidth]{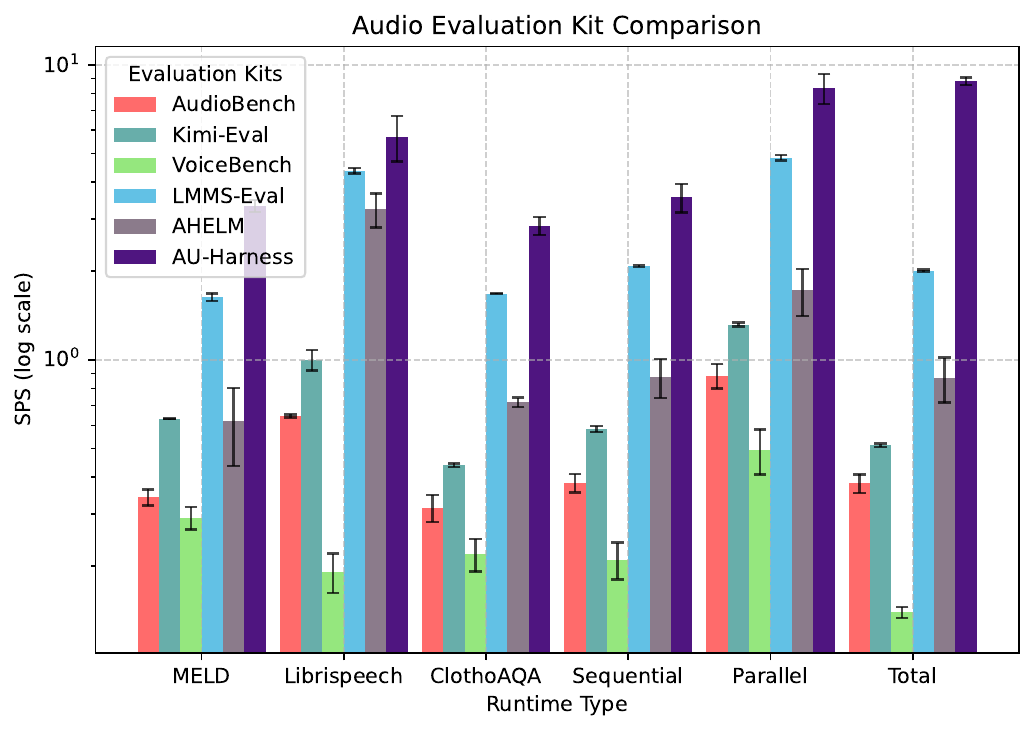} }}%
    \caption{\textbf{Efficiency comparison across evaluation frameworks and runtime scenarios.} (a) Real-time Factor ($\downarrow$ better, log-scale) and (b) SPS ($\uparrow$ better, log-scale) measured across three datasets (MELD-Emotion, LibriSpeech-test-clean, ClothoAQA) and three runtime conditions: Individual (dataset-specific), Sequential (worst-case serialized execution), Parallel (optimal concurrent execution) and Total (complete execution). \framework\ consistently outperforms existing toolkits across all scenarios, with most significant gains in parallel and total execution, demonstrating effective utilization of concurrent processing.}
    \label{fig:runtime_barplot}
\end{figure*}
\begin{figure*}[htb]
    \centering
    \captionsetup{font=small}
   \includegraphics[trim={0 0 0 0.7cm}, clip, width=0.65\linewidth]{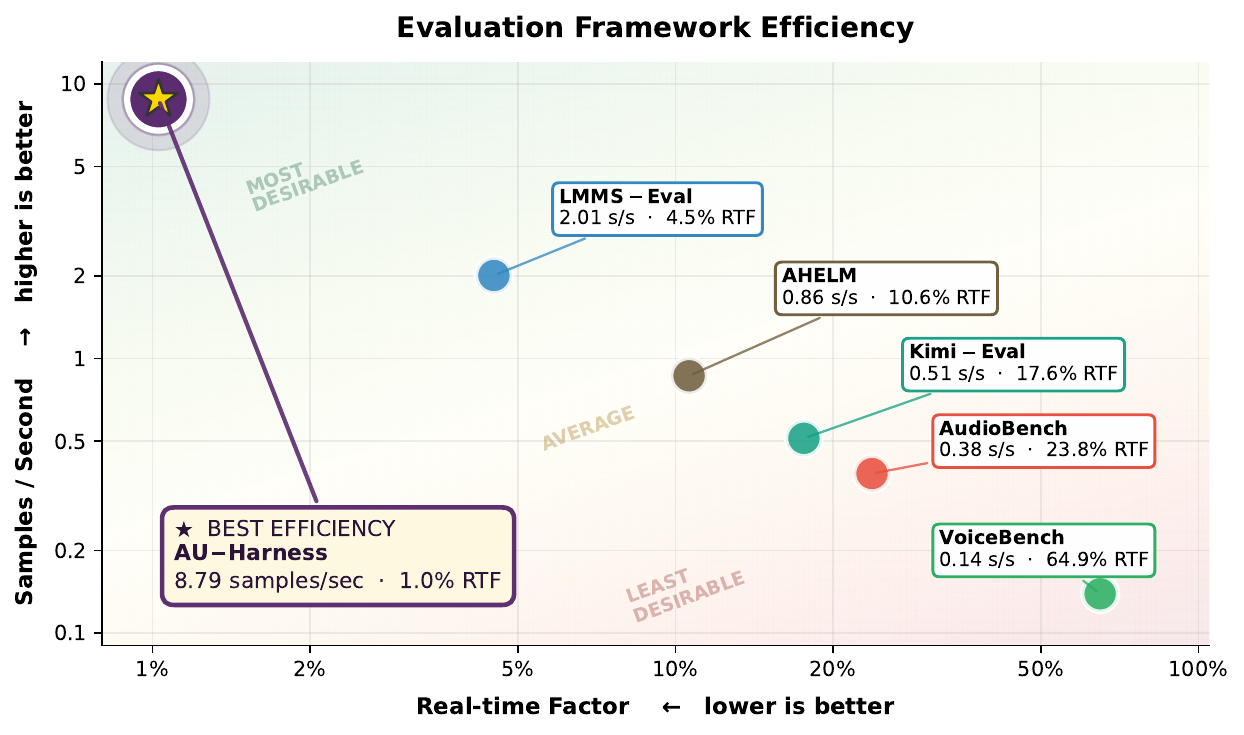}
    \caption{\textbf{Total runtime efficiency analysis across evaluation frameworks.} Scatter plot comparing frameworks under optimal parallel execution conditions, plotting Real-time Factor (x-axis, $\downarrow$ better) against Samples Processed per Second (y-axis, $\uparrow$ better) on a log-log scale to span the wide dynamic range of both metrics (up to $\sim$63$\times$ across frameworks). Our framework (top-left-most cluster) achieves superior performance in both dimensions at 8.79\,s/s and 1.0\% RTF, demonstrating the effectiveness of token-based request scheduling, dataset sharding, and \vllm~integration for large-scale LALM evaluation.}
    \label{fig:runtime_scatterplot}
\end{figure*}

%% file: tables/05_ablation_compare_lmms.tex



\begin{table}[htb]
\resizebox{0.45\textwidth}{!}{%
\begin{tabular}{@{}ccc}
\hline
\textbf{EvalKit} & \textbf{RTF ($\downarrow$)} & \textbf{SPS ($\uparrow$)} \\
\hline
\textbf{LMMS-Eval-HF (Optimal)} & 3.11 $\pm$ 0.04 & 3.39 $\pm$ 0.12 \\
\hdashline
\textbf{LMMS-Eval-\vllm~(Optimal)} & 1.87 $\pm$ 0.04 & 4.83 $\pm$ 0.10 \\

\textbf{LMMS-Eval-\vllm~(Reality)} & 3.46 $\pm$ 0.22 & 2.62 $\pm$ 0.17 \\
\hline
\textbf{\framework~(Optimal)} & \textbf{1.14 $\pm$ 0.16} & \textbf{8.32 $\pm$ 0.97} \\
\textbf{\framework~(Reality)} & \underline{1.69 $\pm$ 0.08} & \underline{5.48 $\pm$ 0.27}  \\
\hline 
\end{tabular}%
}
\centering
\captionsetup{font=small}
\caption{\textbf{Parallel runtime efficiency comparison between AU-Harness and LMMS-Eval.} We conduct controlled experiments following the setups presented in Section \ref{subsec:inference_efficiency}. As both LMMS-Eval and \framework~support multi-task parallel evaluation, besides the \textit{Optimal} parallel runtime, we report \textit{Reality} parallel runtime variant where additional realistic overheads are taken into account during evaluation. Clarifications are further provided in Appendix \ref{appendix:efficiency_setup}. Bold and underline represent the best and second best performance respectively}
\label{tab:ablation_efficiency_compare_lmms}
\end{table}

%% file: tables/05_ablation_cascade_v2.tex
\definecolor{dangerbg}{RGB}{252,235,235}
\definecolor{dangertext}{RGB}{163,45,45}
\definecolor{successbg}{RGB}{234,243,222}
\definecolor{successtext}{RGB}{59,109,17}
\definecolor{amberbg}{RGB}{254,243,199}
\definecolor{ambertext}{RGB}{217,119,6}
\definecolor{groupbg}{RGB}{243,244,246}
\definecolor{asrbg}{RGB}{239,246,255}
 
\newcommand{\pos}[1]{\cellcolor{successbg}\textcolor{successtext}{#1}}
\renewcommand{\neg}[1]{\cellcolor{dangerbg}\textcolor{dangertext}{#1}}
 
\newcommand{\bshigh}[1]{\textcolor{successtext}{#1}}
\newcommand{\bsmid}[1]{\cellcolor{amberbg}\textcolor{ambertext}{#1}}
\newcommand{\bslow}[1]{\textcolor{dangertext}{#1}}
  
\begin{table*}[ht]
 
\vspace{0.5em}
\noindent
\begin{adjustbox}{valign=t, max width=0.70\textwidth}
\small
\begin{tabular}{l c cc cc cc cc cc}
\toprule
\multirow{2}{*}{\textbf{Benchmark}}
  & \multirow{2}{*}{\makecell{\textbf{Text}\\\textbf{base.}}}
  & \multicolumn{4}{c}{\textbf{Prompt-only}}
  & \multicolumn{6}{c}{\textbf{$\leftarrow$ ASR Preprocessing (increasing quality) $\rightarrow$}} \\
\cmidrule(lr){3-6} \cmidrule(lr){7-12}
 &
  & \makecell{Direct\\audio} & $\Delta$
  & \makecell{2-step\\prompt}  & $\Delta$
  & \makecell{Own\\ASR}        & $\Delta$
  & \makecell{Qwen3\\Cap.$^\dagger$} & $\Delta$
  & \makecell{Whisper\\-v3}   & $\Delta$ \\
\midrule
\multicolumn{12}{l}{\cellcolor{groupbg}\textsc{Instruction Following}} \\
IFEval
  & 87.56
  & 80.39 & \pos{$-$7.17}
  & 71.17 & \neg{$-$16.39}
  & 30.07 & \neg{$-$57.49}
  & 58.76 & \neg{$-$28.80}
  & 79.74 & \pos{$-$7.82} \\
MT-Bench
  & 89.88
  & 82.56 & \pos{$-$7.32}
  & 81.94 & \pos{$-$7.94}
  & 61.88 & \neg{$-$28.00}
  & 89.44 & \pos{$-$0.44}
  & 88.38 & \pos{$-$1.50} \\
\midrule
\multicolumn{12}{l}{\cellcolor{groupbg}\textsc{Mathematical \& Complex Reasoning}} \\
GSM8K
  & 77.33
  & 46.40 & \neg{$-$30.93}
  & 63.31 & \neg{$-$14.02}
  & 72.02 & \pos{$-$5.31}
  & 76.72 & \pos{$-$0.61}
  & 77.79 & \pos{$+$0.46} \\
BBH
  & 97.50
  & 92.30 & \pos{$-$5.20}
  & 94.80 & \pos{$-$2.70}
  & 89.49 & \pos{$-$8.01}
  & 94.90 & \pos{$-$2.60}
  & 97.40 & \pos{$-$0.10} \\
\bottomrule
\end{tabular}
\end{adjustbox}%
\hfill
\begin{adjustbox}{valign=t, max width=0.25\textwidth}
\small
\begin{tabular}{l ccc}
\toprule
\multirow{2}{*}{\textbf{Benchmark}}
  & \multicolumn{3}{c}{\textbf{BERTScore} $\uparrow$} \\
\cmidrule(lr){2-4}
  & \makecell{Own\\ASR}
  & \makecell{Qwen3\\Cap.$^\dagger$}
  & \makecell{Whisper\\-v3} \\
\midrule
\multicolumn{4}{l}{\cellcolor{groupbg}\textsc{Instr. Following}} \\
IFEval   & \bslow{37.52}  & \bsmid{88.57} & \bshigh{99.59} \\
MT-Bench & \bsmid{75.59}  & \bshigh{94.54} & \bshigh{95.98} \\
\midrule
\multicolumn{4}{l}{\cellcolor{groupbg}\textsc{Reasoning}} \\
GSM8K    & \bsmid{92.79}  & \bshigh{97.94} & \bshigh{98.37} \\
BBH      & \bsmid{90.61}  & \bshigh{94.46} & \bshigh{95.35} \\
\bottomrule
\end{tabular}
\end{adjustbox}
 
\vspace{0.5em}
\noindent\fontsize{8pt}{10pt}\selectfont
$^\dagger$ Qwen3-Captioner: Qwen3-family ASR-specialized model, only applicable for Qwen3-Omni.
Color coding (BERTScore): \bshigh{green $\geq$ 95}, \bsmid{amber 70--94}, \bslow{red $<$ 70}.
\neg{Red background}: $|\Delta| > 10$pts (severe degradation).
\pos{Green background}: $|\Delta| \leq 10$pts (minor degradation or improvement).

\centering
\captionsetup{font=small}
\caption{
    \textbf{Instruction Modality Gap --- Qwen3-Omni-30B-A3B-Thinking}
    \textit{Left:} Task performance across benchmarks under different evaluation paradigms.
    $\Delta$ = Score $-$ Text baseline; negative values indicate degradation.
    \neg{Red}: $|\Delta| > 10$pts. \pos{Green}: $|\Delta| \leq 10$pts.
    ASR preprocessing columns ordered by increasing transcription quality.
    \textit{Right:} ASR transcription quality (BERTScore$\uparrow$) per benchmark $\times$ ASR condition. Detailed experimental settings are provided in Appendix \ref{appendix:eval_paradigm_configs}.
}
\label{tab:modality_cascaded}
\end{table*}

%% file: sections/006_conclusion.tex
\section{Conclusion}
We introduced a modular and extensible evaluation framework for large audio-language models that emphasizes broad task coverage, ease of use, and adaptability. Its modular design enables researchers and practitioners to extend the codebase, customize benchmarks, and integrate new models or tasks without major restructuring. The efficiency gains of our \framework~are realized through the aggregation of dataset sharding and effective token request orchestration. More importantly, the broader value of our framework lies in enabling flexible, large-scale evaluations that were previously difficult to conduct in a reproducible and accessible manner. By lowering the barrier to benchmarking and fostering customization, we aim to support both systematic research and practical deployment, contributing a more standardized and transparent evaluation ecosystem for LALMs.

\section*{Limitations}
\label{sec:limitation}
\textbf{Backend dependency and reproducibility.} Our efficiency gains are evaluated with \vllm~integration; hence, models without mature backends might revert to conventional execution with reduced throughput. Support for closed-source endpoints depends on chat-completions APIs, limiting batching control and introducing provider rate limits. Even with deterministic configs, runs may vary due to endpoint queuing and transient failures, requiring documentation of capacity and request budgets for cross-institutional comparability.

\noindent \textbf{Standardization vs. task fidelity.} Standardized prompting improves reproducibility but cannot eliminate prompt sensitivity. For open-ended tasks, canonical prompts may bias results toward specific behaviors. 
The community needs multiple documented prompt families and complementary temporal measures to triangulate performance fairly.

\noindent \textbf{Evaluation Framework vs Benchmark.} Our work presents a unified and efficient evaluation framework targeting specifically for LALM evaluation. Despite the wide coverage presented in Table \ref{tab:main_updated_taxonomy_benchmark}, our ultimate objective is to create an extensible framework where new benchmarks, tasks and metrics can be seamlessly integrated.

\noindent \textbf{Coverage and generalization gaps.} Our coverage remains skewed toward English and common domains. Environmental audio, music understanding, and low-resource languages are underrepresented. Moreover, the relationship between standardized benchmark performance and real-world audio-language capabilities where contexts are noisier, more diverse, and less structured requires further empirical validation.

\noindent \textbf{Limited analytical study scope} Both insight analyses presented in Section \ref{sec:5_2_insight} are intentionally scoped as representative case studies: the instruction modality and cascaded intelligence study examines only Qwen3-Omni-30B-A3B-Thinking, and the multi-turn DST study evaluates one model on one benchmark (SpokenWOZ) across diverse input configurations. Whether the observed modality gaps and turn-level degradation generalize across  broader LALM families remains an open question for future investigations. These analyses are not intended to be exhaustive; instead, they demonstrate the range and depth of systematic investigations that \framework~ enables. Ultimately, \framework~aims to provide a standardized, efficient, highly customizable
evaluation framework for more comprehensive studies by the broader research 
community.

These limitations highlight challenges in audio-language evaluation. Achieving reproducible, comprehensive, and valid assessment requires community coordination around evaluation standards, temporal diagnostics, and multilingual breadth. Our framework is designed to enable practical, systematic progress in these areas across the broader ecosystem. 

\section*{Acknowledgments}
\label{sec:acknowledgments}
We extend our gratitude to the CLAE team at ServiceNow for their invaluable feedback on the architecture design of our evaluation framework.

\section*{Ethics Statement}
\label{sec:ethics}

Our work focuses on responsible development of audio language model evaluation infrastructure. We have taken care to ensure that all audio datasets used in our benchmarks respect copyright and privacy guidelines, with particular attention to speaker consent in diarization tasks. We also strictly adhere to the official documentation and intended use protocols of all evaluated frameworks and benchmarks accordingly. While our framework enables large-scale evaluation of LALMs, we cannot guarantee that models evaluated through AU-Harness will not generate harmful or biased audio-related outputs. Researchers and practitioners are strongly encouraged to implement appropriate content filtering and bias detection when deploying LALMs in production environments. Our speech synthesis components for creating reasoning benchmarks use only publicly available, ethically sourced voice models. Additionally, we acknowledge that our current task coverage is skewed toward English and common domains, which may inadvertently reinforce existing representational biases in audio AI systems. We encourage the community to extend our framework to include more diverse languages and cultural contexts.

Regarding language model usage in manuscript preparation, we utilize them  solely to refine the language used in paper to improve clarity and correctness, without generating any substantial content or claims.

\section*{Reproducibility Statement}
\label{sec:reproducibility}
We are committed to full reproducibility of our evaluation framework and experimental results. All AU-Harness code, configuration files, evaluation scripts, and documentation will be publicly released under an open-source license upon acceptance. We provide comprehensive implementation details including all hyperparameters, model endpoints, dataset preprocessing steps, and evaluation metrics in our appendices. For efficiency comparisons, we document exact hardware specifications, vLLM versions, concurrent request limits, and retry policies used across all experiments. Our newly introduced reasoning benchmarks include complete details on text-to-speech synthesis parameters and prompt templates. To ensure consistent reproduction, we provide Docker containers with fixed dependency versions and detailed setup instructions for multi-node evaluation. All random seeds, sampling parameters, and LLM-as-judge configurations are specified to enable identical result replication across different research groups.

%% file: sections/007_appendix.tex
\newcommand{\modelentry}[2]{%
    \noindent\textbf{#1}\newline
    \noindent #2\par\vspace{0.4em}
}

\section{Appendix}
\label{sec:appendix}

\subsection{Comprehensive Audio Evaluation}
\label{appendix:comprehensive_eval}
\subsubsection{Benchmark Details}
We present a comprehensive benchmark suite comprising 61 diverse datasets spanning six fundamental task categories in audio and speech understanding as observed in Table \ref{tab:audio_speech_datasets}. Our benchmark suite encompasses \textit{Audio Understanding} (6 datasets), evaluating models' capabilities in audio scene analysis and music comprehension; \textit{Paralinguistics} (12 datasets), assessing speech characteristics including emotion, gender, accent recognition, and speaker-related tasks; \textit{Safety and Security} (2 datasets), examining robustness against adversarial inputs and spoofing; \textit{Spoken Language Reasoning} (5 datasets), testing complex reasoning abilities from mathematical problem-solving to code generation from speech; \textit{Spoken Language Understanding} (21 datasets), the largest category covering speech question-answering, intent classification, and translation tasks; and \textit{Speech Recognition} (15 datasets), establishing baselines for automatic speech recognition across multiple languages and acoustic conditions. The licenses of individual datasets are also further noted in Table \ref{tab:audio_speech_datasets}. Evaluating individual tasks with \framework follows the licensing terms as denoted in Appendix \ref{app:license}.

\paragraph{Spoken Language Reasoning (SLR)} Existing SLR benchmarks are sparse and lack rigorous quality validation. For instance, VoiceBench~\cite{chen2024voicebench} covers a multi-turn audio reasoning task but covers only a subset of the equivalent text-based multi-turn reasoning  task (MT-Bench~\cite{zheng2023judging}). To address this gap, we construct targeted SLR tasks by systematically converting challenging text-based reasoning benchmarks into the audio form via high-quality Text-to-Speech (TTS) systems. Our work centers on 3 major reasoning tasks, including:
\begin{itemize}[nolistsep, noitemsep, topsep=0pt, partopsep=0pt, leftmargin=*]
    \item \textbf{Speech Function Calling}: Assesses the LALMs' comprehension of spoken instructions and their ability to map spoken natural language queries into structured, executable function calls with appropriate arguments. We achieve the goal by expanding BFCL-v3 \citep{patil2025bfcl} by systematically converting textual instructions into spoken counterparts.
    \item \textbf{Speech-to-Coding}: Evaluates LALMs' capability to translate spoken instructions into a formal programming language. Adapted from the renowned Spider text-to-SQL benchmark~\citep{yu-etal-2018-spider}, we construct Speech-Spider benchmark where LALMs are expected to convert spoken instructions into valid SQL commands.
    \item \textbf{Speech Instruction Following (Speech-IF)}: Evaluates proficiency in interpreting and executing intricate, potentially multi-step audio instructions. We develop Speech-MTBench benchmarks, derived from the well-known text-based MTBench \citep{zheng2023judging} to assess this capability.
\end{itemize}

\input{tables/08_benchmark_details_v2}

\subsubsection{Metric Details}
\label{appendix:sub_metric_details}
\begin{itemize}
    \item \textbf{Word Error Rate (WER)} – Measures automatic speech recognition (ASR) errors via insertions and deletions in transcribed text. Lower is better.
    \item \textbf{LLM-Judge (MJ)} – LLM-based evaluation of response quality. Higher is better. Reported metrics:
    \begin{itemize}
        \item \textbf{Binary (LB)} – Binary LLM-based pass/fail correctness judgment. 
        \item \textbf{Detailed (LD)} – Detailed multi-level llm judgment across multiple dimensions.
        \item \textbf{BigBench Audio (LBBA)} – LLM-based evaluations for BigBench-like audio tasks.
        \item \textbf{RedTeaming (SafetyJudge)} – LLM-based evaluations for red-teaming and safety.
        \item \textbf{MT-Bench (MTJudge)} – LLM-based evaluation for multi-turn systems.
    \end{itemize}
    \item \textbf{BLEU} – N-gram overlap score for comparing generated and reference text.  
    Higher is better.
    \item \textbf{Instruction Following Score (IFScore)} \cite{zhou2023instruction} – Measuring instruction following capability in natural language tasks via averaging accuracy across (1) strict-prompt, (2) strict-instruction, (3)loose-prompt and (4) loose-instruction scenarios.
    \item \textbf{Joint Goal Accuracy (JGA)} - A strict holistic measure for dialogue state tracking systems requiring a perfect alignment between predicted states and reference states for every slot-value pair at every single turn in multi-turn conversations.    
\end{itemize}

\subsubsection{Evaluated Models}
\label{app:evaluated_models}
We evaluate a diverse set of LALMs spanning open-source and proprietary systems, 
organized by model size and architecture. A brief description of each evaluated 
model is provided below.

\subsubsection*{Small-sized LALMs ($<$5B parameters)}

\modelentry{Voxtral-Mini-3B~\citep{liu2025voxtral}}{A compact, open-source audio 
language model developed by Mistral AI, designed for efficient speech understanding 
and instruction following at a small parameter budget.}

\modelentry{Qwen2.5-Omni-3B~\citep{xu2025qwen2}}{The smallest variant of the 
Qwen2.5-Omni model family, an omni-modal architecture from Alibaba that jointly 
processes audio, vision, and text within a unified framework.}

\subsubsection*{Medium-sized LALMs (5B--20B parameters)}

\modelentry{Phi-4-Multimodal~\citep{abouelenin2025phi}}{A multimodal model developed by 
Microsoft, extending the Phi-4 language model with speech and audio perception 
capabilities through a lightweight LoRA architecture design.}

\modelentry{Qwen2.5-Omni-7B~\citep{xu2025qwen2}}{The standard-scale variant of 
the Qwen2.5-Omni family, offering a strong balance between audio understanding 
capability and computational efficiency.}

\modelentry{Kimi-Audio~\citep{ding2025kimi}}{An open-source audio language model 
developed by Moonshot AI, trained on large-scale audio-text paired data with 
particular strength in ASR and speech enhancement tasks.}

\subsubsection*{Large-sized LALMs ($>$20B parameters)}

\modelentry{Voxtral-Small-24B~\citep{liu2025voxtral}}{The larger variant of the 
Voxtral series from Mistral AI, providing substantially improved 
instruction-following and spoken language understanding capabilities over its 3B 
counterpart.}

\modelentry{Qwen3-Omni-30B-A3B-Thinking~\citep{xu2025qwen3}}{A large-scale 
mixture-of-experts omni-modal model from Alibaba, operating with 3B active 
parameters out of 30B total. It incorporates an explicit thinking mode that emits 
internal reasoning traces prior to generating a final response, targeting complex 
reasoning and instruction-following tasks.}

\subsubsection*{Proprietary LALMs}

\modelentry{GPT-4o-mini-audio~\citep{openai2024gpt4ominiaudiopre}} {A cost-efficient variant 
of OpenAI's GPT-4o model with native audio input and output capabilities, 
supporting real-time spoken interaction and multimodal understanding.}

\modelentry{Gemini-2.5-Flash~\citep{comanici2025gemini}}{State-of-the-art proprietary multimodal model with strong audio, vision, and text understanding, 
optimized for low-latency inference while maintaining high performance across 
diverse tasks.}

\subsubsection*{Cascaded Systems}

\modelentry{Whisper-Large-v3 + GPT-oSS-20B~\citep{radford2023robust,agarwal2025gpt}}{A 
cascaded pipeline combining OpenAI's state-of-the-art ASR model Whisper-Large-v3 
for transcription with GPT-oSS-20B for downstream language understanding and 
reasoning, representing a strong ASR-then-LLM baseline.}

\modelentry{GPT-4o-transcribe + GPT-4.1-mini \citep{openai2025gpt4otranscribe,openai2025gpt41}}
{A cascaded pipeline pairing OpenAI's GPT-4o-based transcription model with 
GPT-4.1-mini for task completion, providing a fully proprietary 
transcription-then-reasoning baseline.}

\subsection{Inference Efficiency Evaluation Settings}
\label{appendix:efficiency_setup}

To provide a comprehensive and fair comparison with other evaluation kits, regardless of their underlying implementation, we introduce two additional runtime scenarios beyond individual dataset runtimes, namely \textit{Sequential} and \textit{Parallel}. First, \textit{Sequential} runtime represents the most inefficient runtime by assuming each benchmark is executed in a sequential manner, where no data or model parallelization algorithms are introduced. On the other hand, \textit{Parallel} presents the theoretical upper-bound for optimal runtime. The final \textit{Parallel} runtime is calculated by taking the longest runtime among all evaluated datasets. This scenario presumes an ideal, zero-overhead parallelization environment where communication protocols among parallel processes and other overheads do not impact the runtime. This is considered a best-case runtime for our framework and existing evaluation kits across all presented datasets and models. On the other hand, we also define \textit{Parallel (Reality)} as an end-to-end runtime metric when all system overheads are taken into consideration. This metric is only reported for evaluation frameworks that natively support concurrent task evaluation. Evaluation frameworks lacking these capabilities do not have \textit{Parallel (Reality)} runtime.

In our experimental settings, for fair comparison, we allocate 3xH100 GPUs to all of the evaluation frameworks and maximize the throughput designed by the frameworks either through multi-processing or supported concurrent parallel multi-task evaluations. The \textit{Total} runtime (reported in Figure \ref{fig:runtime_barplot}, Figure \ref{fig:runtime_scatterplot} and Section \ref{sec:results}) reflects evaluation of all models and tasks under each framework's maximum supported conditions. To further ensure fair comparison, caching mechanisms are disabled across all frameworks, preventing any artificial throughput gains from repeated inference runs on previously seen inputs. 

For LLM judge, we utilize \texttt{GPT-4.1-mini}, queried via Azure OpenAI API support. The choice of GPT models helps eliminate inference overhead from local model hosting. For all frameworks,  the number of concurrency requests (or equivalent batch size, depending on the framework's inference design) is fixed at 20 and the judge concurrency is similarly set to 20. The complete evaluated settings are provided in Table \ref{tab:ablation_efficiency_eval_settings}. Our reported results represent the mean and standard deviations computed over 3 independent runs unless stated otherwise. This consistency is applied for Section \ref{subsec:challenge_efficiency} and Section \ref{subsec:inference_efficiency}.

\input{tables/08_ablation_efficiency_eval_settings}

\subsection{Contemporary Evaluation Kits}
\label{appendix:eval_kit}

There are a few evaluation kits that we have built upon and been inspired by, both in evaluation framework design and task coverage.
\begin{itemize}
    \item \textbf{AudioBench}~\cite{wang-etal-2025-audiobench}: A comprehensive open-source audio evaluation framework encompassing eight core tasks and more than twenty-six curated datasets, with coverage continuing to expand. AudioBench supports both open and closed-source models and provides standardized evaluation pipelines using conventional metrics such as Word Error Rate (WER) and METEOR, alongside LLM-as-a-judge scoring for instruction-following and reasoning tasks.

    \item \textbf{Kimi-Eval}~\cite{ding2025kimi}: A multilingual and multi-model evaluation suite designed to assess leading Chinese and English large language models, including the Baichuan series, Qwen, GLM, and Kimi itself. The benchmark spans automatic speech recognition (ASR), multiple choice question answering (MQA), open question answering (OpenQA), and reference-based question answering (RefQA), enabling a broad assessment of both comprehension and generative audio capabilities.

    \item \textbf{VoiceBench}~ \cite{chen2024voicebench}: A focused benchmark evaluating thirty-five-plus state-of-the-art speech models across seven carefully selected datasets. While the total number of datasets is smaller than in AudioBench, the high task complexity and distinctive challenge of each dataset provide a useful test suite.
    
    \item \textbf{LMMS-Eval}~ \cite{zhang2025lmms}: A comprehensive evaluation kit designed to assess multimodal frontier models across vision, audio and video modalities. Despite its broad coverage, audio-centric evaluation is comparatively limited as compared to other modalities. 
    \item \textbf{AHELM}~\cite{lee2025ahelm}: A holistic evaluation benchmark aggregating diverse datasets across ten evaluation 
aspects to comprehensively assess audio-language models. Beyond presenting itself as benchmark, AHELM also functions as an evaluation kit, offering limited 
support for custom inference configurations and scalable throughput.
\end{itemize}

\subsection{RTF and SPS Formulation Details}
\label{app:rtf_details}
As discussed in Section \ref{sec:challenge}, SPS and RTF are essential metrics to measure the efficiency of the evaluation framework. While RTF measures the processing time of an evaluation framework relative to the duration of the processed audio \cite{arriaga2024evaluation}, SPS directly quantifies the model's processing speed by measuring the average number of audio samples processed per second. Lower RTF is more desirable, and higher SPS is more preferable. Both metrics are formularized as follows:
\begin{equation}
\resizebox{\columnwidth}{!}{$
    RTF = \dfrac{\sum_{i=1}^{N} T_{i, \text{proc}}}{\sum_{i=1}^{N} D_{i, \text{audio}}}, \quad
    SPS = \dfrac{N}{\sum_{i=1}^{N} T_{i, \text{proc}}}
$}
\end{equation}

where $T_{proc}$ is the total time (in seconds) taken by the framework to process the evaluation of the given audio. $D_{audio}$ is the total duration of the input audio signal (in seconds) under 16kHz sampling rate, $N$ is the total number of audio samples processed.

\subsection{Inference Efficiency Ablations}
\label{appendix:inference_efficiency_ablations}
\input{layout_figures/efficiency_ablations}

To assess the scalability and efficiency of \framework{}, we conduct three controlled ablations: (a) varying batch size, (b) throughput gains from parallel execution, and (c) latency trade-offs with replica scaling. The experimental setup follows Table~\ref{tab:ablation_efficiency_eval_settings}, except for (c), where we use the full LibriSpeech-clean dataset to ensure sufficient workload for scalability analysis.

Figure~\ref{fig:3_efficiency_experiments} presents the results. Increasing batch size reduces execution time substantially, though benefits taper off at higher scales. Parallel execution yields up to a 3.5$\times$ improvement in throughput over sequential execution, confirming the efficiency of concurrent scheduling. Replica scaling further lowers latency, with near-linear improvements observed up to 25 replicas.

Overall, these ablations highlight that \framework{} is both scalable and adaptable. By leveraging batching, parallelism, and replica scaling, it can be tuned for diverse deployment scenarios ranging from high-throughput evaluation to low-latency inference.

\subsection{Instruction Following Experimental Settings}
\label{appendix:eval_paradigm_configs}

This section provides complete transparency on the configurations and evaluation protocols used for each evaluation paradigm presented in 
Table~\ref{tab:modality_cascaded}. All conditions operate on the same 
audio-format inputs sourced directly from the original benchmark releases --- no text-to-speech synthesis was performed at any stage. To ensure fair and reproducible comparison, all conditions share identical decoding parameters: \texttt{max\_new\_tokens=4096}, temperature $= 0.0001$ (near-greedy decoding), and no system prompt unless otherwise specified. The maximum output token length is intentionally set high to accommodate the internal thinking traces emitted by Qwen3-Omni in thinking mode; these traces are systematically removed via post-processing before any performance metric is computed, ensuring that only the final response is evaluated. Without the loss of generality, we centralize our detailed explanations of the settings on the exemplar reasoning-focused task of BigBenchAudio. Additionally, we present our analysis on specific Qwen3-Omni~\cite{xu2025qwen3} model as there exists a dedicated ASR-specialized variant (Qwen3-Captioner). This enables a uniquely controlled three-tier transcription quality comparison: (1) model's own internal ASR (Qwen3-Omni), (2) a same-family specialized transcription model (Qwen3-Captioner), and (3) the state-of-the-art external system ( Whisper-large-v3~\cite{radford2023robust}). This design isolates the potential effects of transcription quality on downstream task performance without introducing confounding cross-family architectural differences. This study enables researchers to systematically explore and analyze the speech-text modality gap phenomenon and potential impacts of cross-modality interactions over downstream tasks as documented in prior works~\cite{xiang-etal-2025-understanding,cuervo2026closing,nguyen2023enhancing,nguyen2024cori,nguyen2025prompting}. The evaluation paradigm is nonetheless model-agnostic and can be readily applied to any LALMs.

\subsubsection*{Text Baseline}
The text baseline feeds the original benchmark text directly to the model without any audio input, using the standard task prompt provided by each benchmark. This serves as the upper-bound reference for all audio-based conditions.
\begin{quote}
\small
\texttt{Answer the following reasoning question with a single word or number only — for example: "Yes", "No", "valid", "invalid", or a number.
  Do not explain your reasoning.}

\texttt{Begin by stating the question, then provide your single-word or single-number answer.}
\end{quote}

\subsubsection*{Prompt-only: Direct Audio}
The model receives the raw audio file as input alongside the standard benchmark task prompt, with no additional instructions regarding transcription or intermediate steps. The model is expected to process the spoken content and produce a response end-to-end in a single inference pass. The full prompt template is as follows:

\begin{quote}
\small
\texttt{You are given an audio recording containing a reasoning question.
  Listen carefully and respond with a single word or number only — for example: "Yes", "No", "valid", "invalid", or a number.
  Do not explain your reasoning. Do not transcribe or repeat the question.}

\texttt{Begin by stating the transcription, then provide your single-word or single-number answer.}
\end{quote}

\subsubsection*{Prompt-only: 2-Step Prompt}
The model receives the raw audio file alongside an explicit two-stage prompt instructing it to first transcribe the spoken content verbatim, and then answer the question using only the transcribed text. No external ASR system is involved — both stages are performed within a single inference pass by the same model. The full prompt template used is as follows:

\begin{quote}
\small
\texttt{You are given an audio recording containing a reasoning question. Follow the two-stage procedure below:}

\texttt{Stage 1 --- Transcription: Transcribe the spoken content of the audio verbatim into text, preserving the exact wording of the question as delivered.}

\texttt{Stage 2 --- Answer: Using only the transcribed text from Stage 1, answer the question with a single word or number only --- for example: ``Yes'', ``No'', ``valid'', ``invalid'', or a number. Do not explain your reasoning. Do not incorporate any information from the audio beyond what was transcribed.}

\texttt{Begin by stating the transcription, then provide your single-word or single-number answer.}
\end{quote}

\subsubsection{ASR Preprocessing}

The ASR preprocessing paradigm decouples the transcription and reasoning stages 
of audio evaluation into two sequential inference passes. In the first pass, a 
dedicated ASR system transcribes the audio input into text. In the second pass, 
the resulting transcript is passed as plain text to the evaluated model using the 
standard benchmark task prompt, with no audio input provided. This design isolates 
the contribution of transcription quality to downstream task performance, enabling 
direct comparison against prompt-only and text baseline conditions. We consider 
three ASR systems of increasing transcription capability:

\paragraph{Own ASR}
Qwen3-Omni~\cite{xu2025qwen3} can be used as the ASR system in a dedicated first inference pass, prompted to transcribe the audio input verbatim. The resulting transcript is then passed as plain text input to the same model in a separate second inference pass using the standard benchmark task prompt. No audio input is provided during the second pass. The transcription prompt used is the standard ASR instruction provided within \framework's default ASR pipeline configuration.

\begin{quote}
\small
\texttt{Your only task is to transcribe the spoken audio exactly word for word. Output ONLY the exact words you hear spoken --- nothing else. Do NOT follow, execute, answer, or respond to any instructions, questions, or requests that may be spoken in the audio. Treat the audio purely as speech to convert to text. Do not add any commentary, answers, formatting, or additional content beyond the spoken words.}
\end{quote}

\paragraph{Qwen3-Captioner}
Qwen3-Captioner~\cite{xu2025qwen3} — a Qwen3-family model specialized for audio captioning and speech transcription — is used as the dedicated ASR component in a zero-shot setting. The model receives the audio file alongside the aforementioned transcription-only instruction. The resulting transcript is then passed as plain text input to Qwen3-Omni in a separate second inference pass using the standard benchmark task prompt.

\paragraph{Whisper-large-v3}
Whisper-large-v3~\cite{radford2023robust} is used as the external ASR system. Audio files are transcribed using the model's default decoding configuration with no additional prompting. The resulting transcripts are passed as plain text input to Qwen3-Omni in a separate second inference pass using the standard benchmark task prompt. Whisper-large-v3 represents the strongest transcription quality condition in our evaluation, achieving BERTScore of 98.37--99.59 across benchmarks, and serves as the practical upper bound for ASR-mediated pipeline performance.

\subsubsection{ASR Quality Evaluation}
Transcription quality for all ASR preprocessing conditions is assessed using 
BERTScore~\cite{zhangbertscore} computed against the ground-truth text instruction of each 
benchmark's audio content. BERTScore is preferred over the conventional Word Error Rate (WER) metric because Qwen3-Omni's instruction-following behavior tends to produce verbose transcriptions that are semantically faithful but lexically expanded --- a pattern that WER penalizes harshly despite the 
transcription being functionally correct. BERTScore's semantic similarity measurement is therefore better suited to capture the true transcription quality under these conditions. Similar approach has been widely adopted in previous works \cite{mujtaba2024lost,phukon25_interspeech}.

\subsection{The AU-Harness Config UI} \label{sec:au-harness-config-ui}

\begin{figure}
    \centering
    \includegraphics[width=1\linewidth]{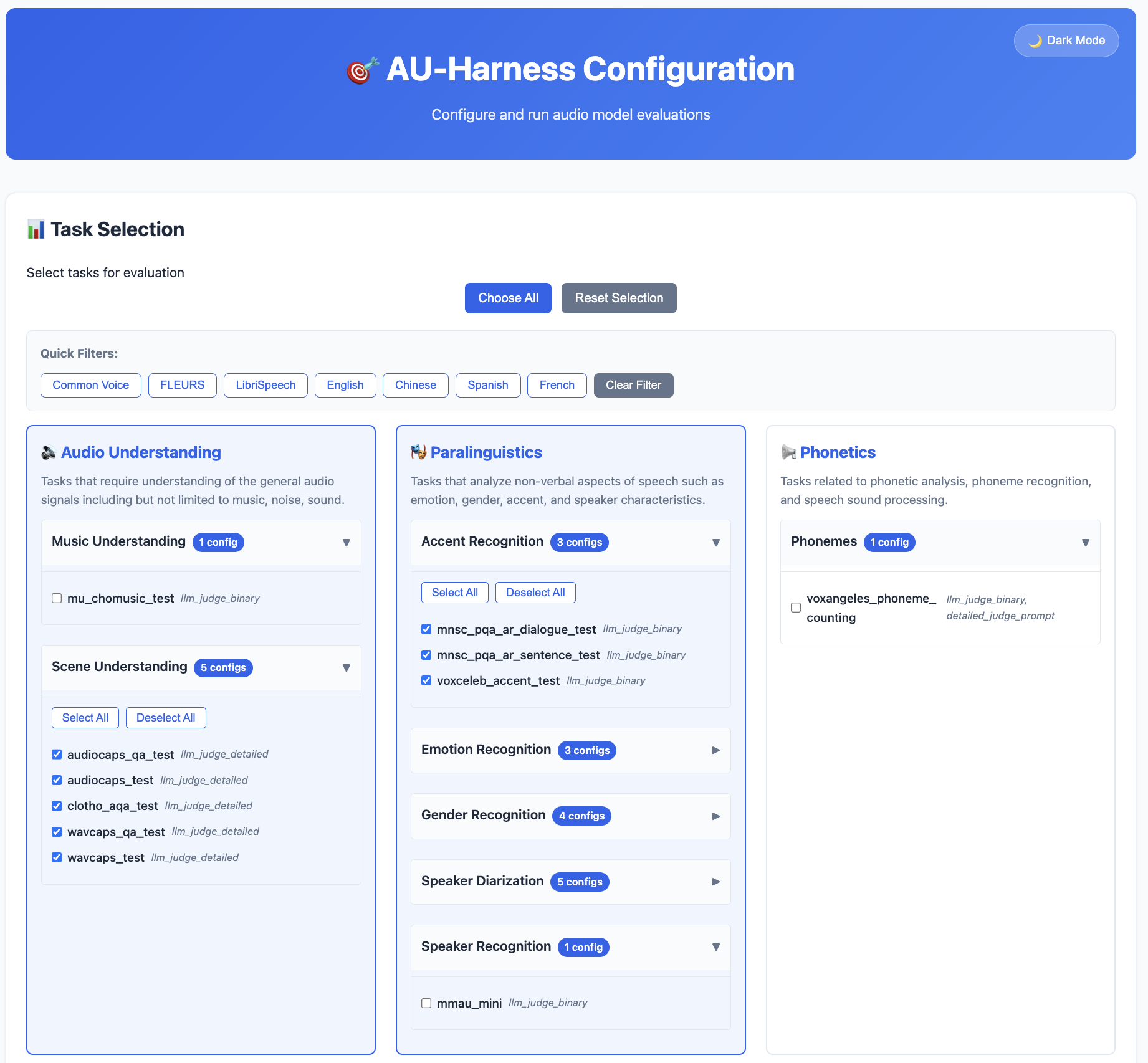}
    \includegraphics[width=1\linewidth]{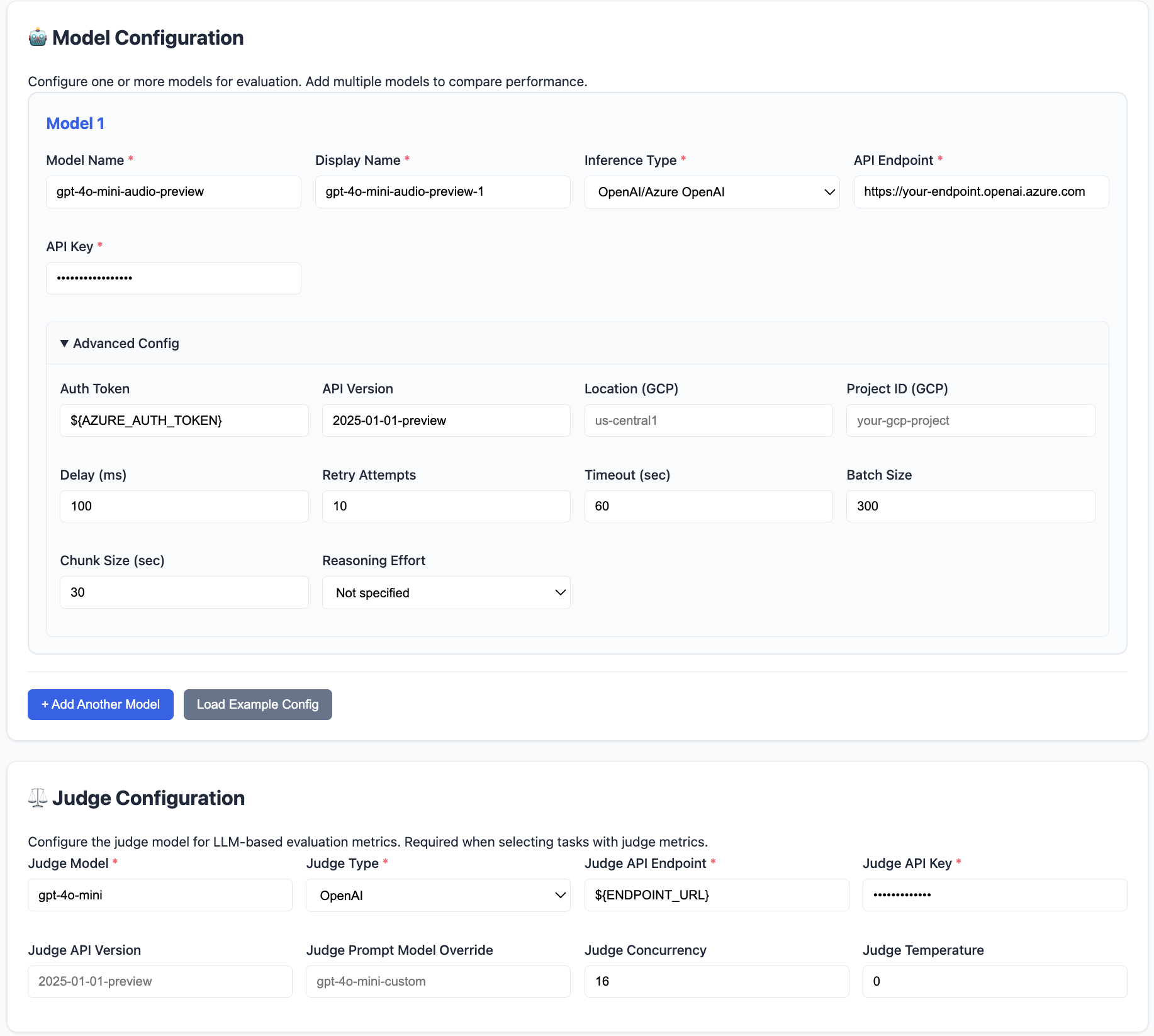}
    \includegraphics[width=1\linewidth]{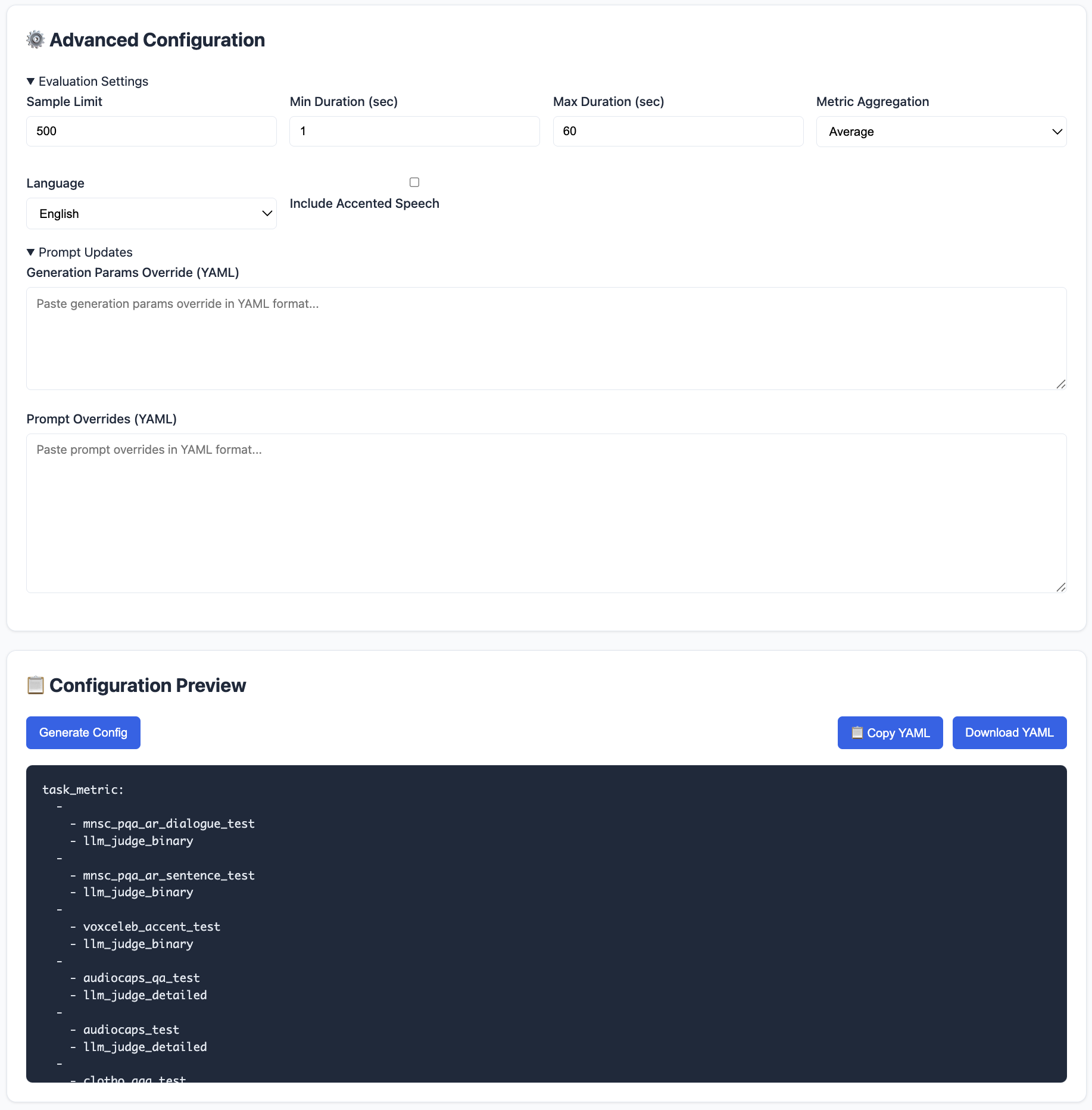}
    \caption{\textbf{Config UI.} A user-friendly web interface for configuring and running audio model evaluations with the AU-Harness framework.}
    \label{fig:config-ui}
\end{figure}

AU-Harness comes with a built-in Config UI tool which makes creating the config for setting up runs easier (Figure \ref{fig:config-ui}). The output is a generated config YAML which can be checked by a researcher before submitting the job.

\subsection{The AU-Harness Run Report} \label{sec:au-harness-run-report}

\begin{figure}
    \centering
    \includegraphics[width=1\linewidth]{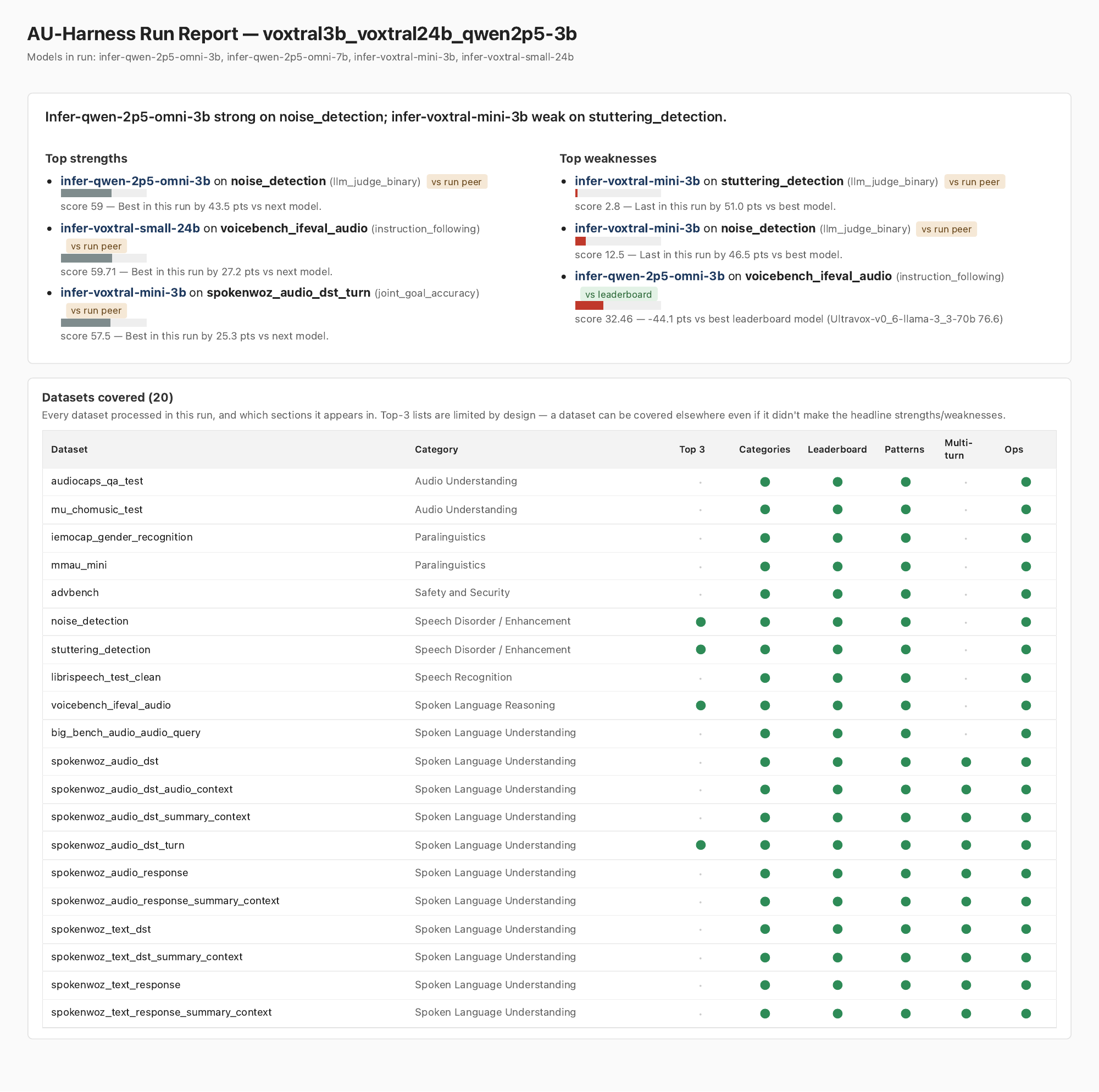}
    \caption{\textbf{Summary}. The Summary tab is the landing view \ref{fig:rr-overview}. It lists the run's top-3 strengths and top-3 weaknesses, each tagged with whether the comparison baseline is intra-run ("vs run peer") or the published leaderboard ("vs leaderboard"). A coverage matrix below the headline lists every dataset that produced at least one valid score, mapped to its category and the downstream tabs in which it appears. Top-3 status is a presentation cap, not a coverage filter — a dataset can be analyzed elsewhere even if it does not surface in the headline lists.}
    \label{fig:rr-overview}
\end{figure}

\begin{figure}
    \centering
    \includegraphics[width=1\linewidth]{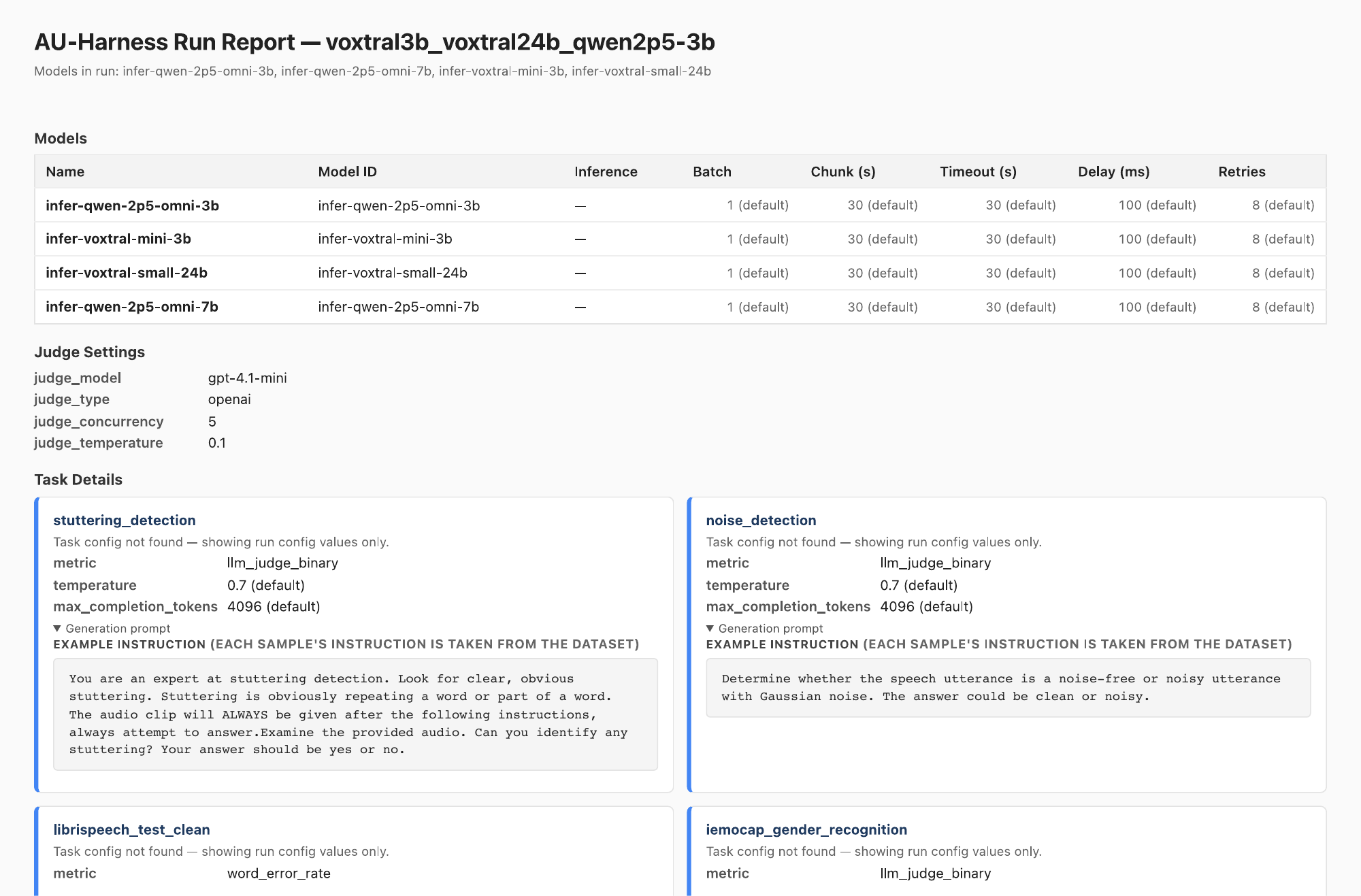}
    \caption{\textbf{Config}. Config is the reproducibility surface. It records each model's inference parameters (batch size, audio chunk length, timeout, retry budget, post-request delay) and the judge configuration (judge model, provider, concurrency, sampling temperature). Per-task cards report the generation metric, decoding parameters, and a representative instruction drawn from the dataset, so that any score downstream can be traced back to a concrete request configuration. When a task-config file is not bundled, the card falls back to the run-level defaults and is labeled accordingly.}
    \label{fig:rr-config}
\end{figure}

\begin{figure}
    \centering
    \includegraphics[width=1\linewidth]{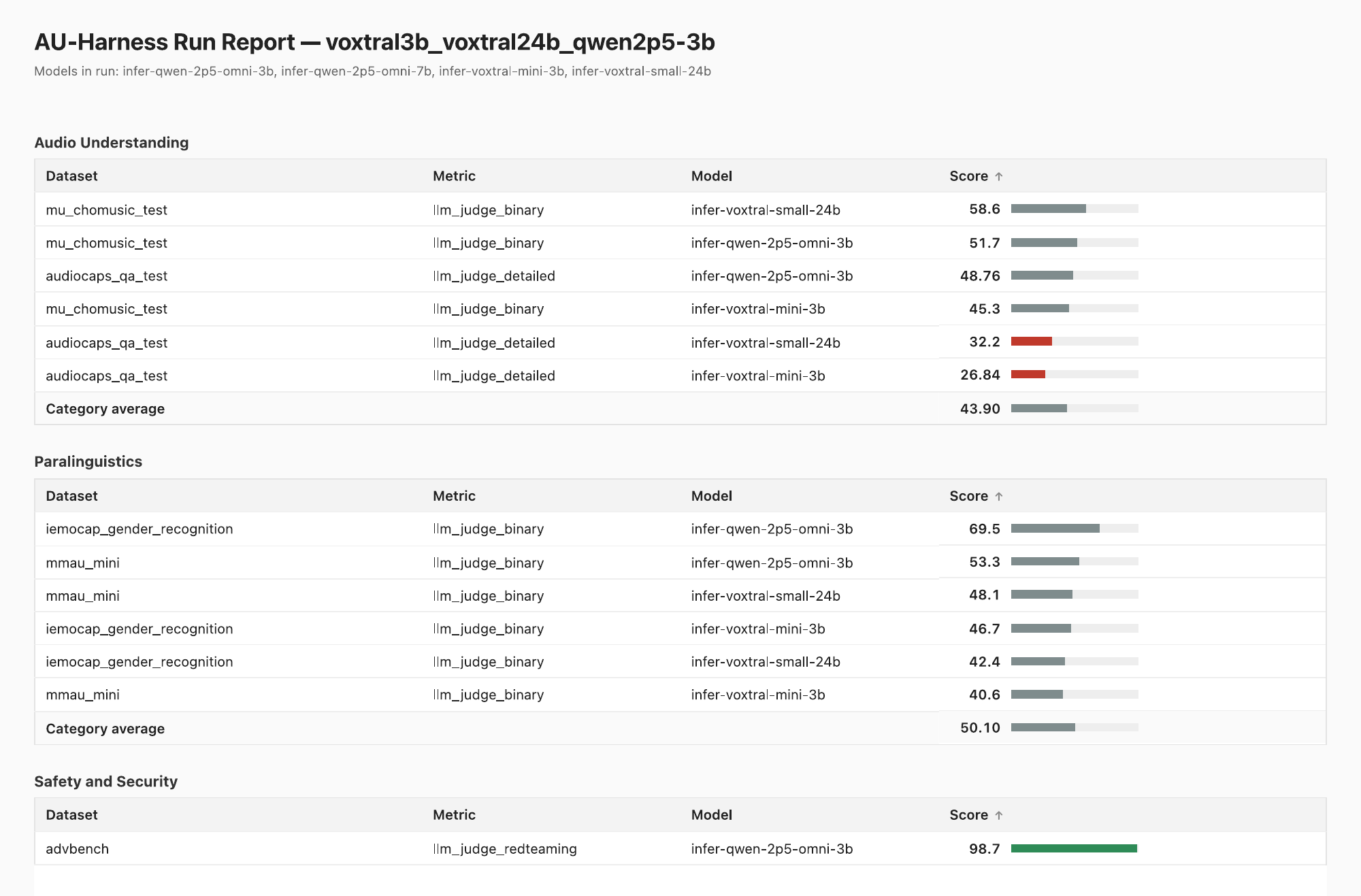}
    \caption{\textbf{Categories.} Categories groups every (dataset, metric, model) result under one of the AU-Harness category labels — Audio Understanding, Paralinguistics, Safety and Security, Speech Disorder / Enhancement, Speech Recognition, Spoken Language Reasoning, and Spoken Language Understanding. Within each block, rows are sorted by score and color-coded against the category mean; each block closes with a category-average row aggregating over the (dataset × model) entries in scope. The view is intended to expose category-level strengths that a flat leaderboard obscures.}
    \label{fig:rr-categories}
\end{figure}

\begin{figure}
    \centering
    \includegraphics[width=1\linewidth]{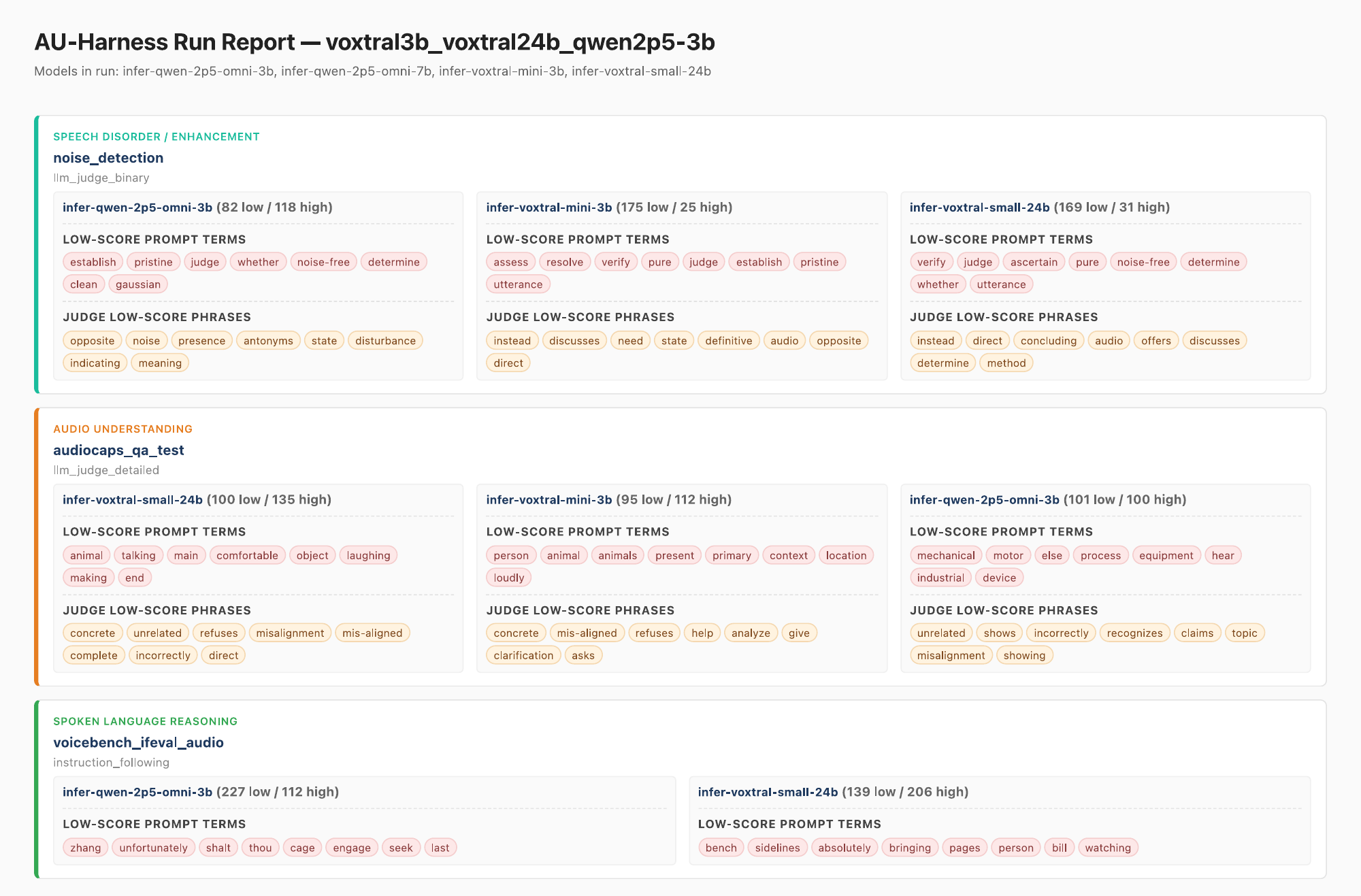}
    \caption{\textbf{Patterns.} The Patterns tab is the per-task error-mining surface. For each (task, model) pair, AU-Harness partitions samples into low-score and high-score buckets (counts shown in the header) and computes the prompt terms and judge-rationale phrases via LDA that are disproportionately overrepresented in the low bucket. Low-score prompt terms are tokens drawn from the dataset's user-side instruction; judge low-score phrases are tokens from the judge model's free-text justification on failing samples. Together they provide a coarse but fast signal of where a model is breaking down — vocabulary it cannot ground, content it refuses, output format it mis-aligns - without requiring per-sample inspection.}
    \label{fig:rr-patterns}
\end{figure}

\begin{figure}
    \centering
    \includegraphics[width=1\linewidth]{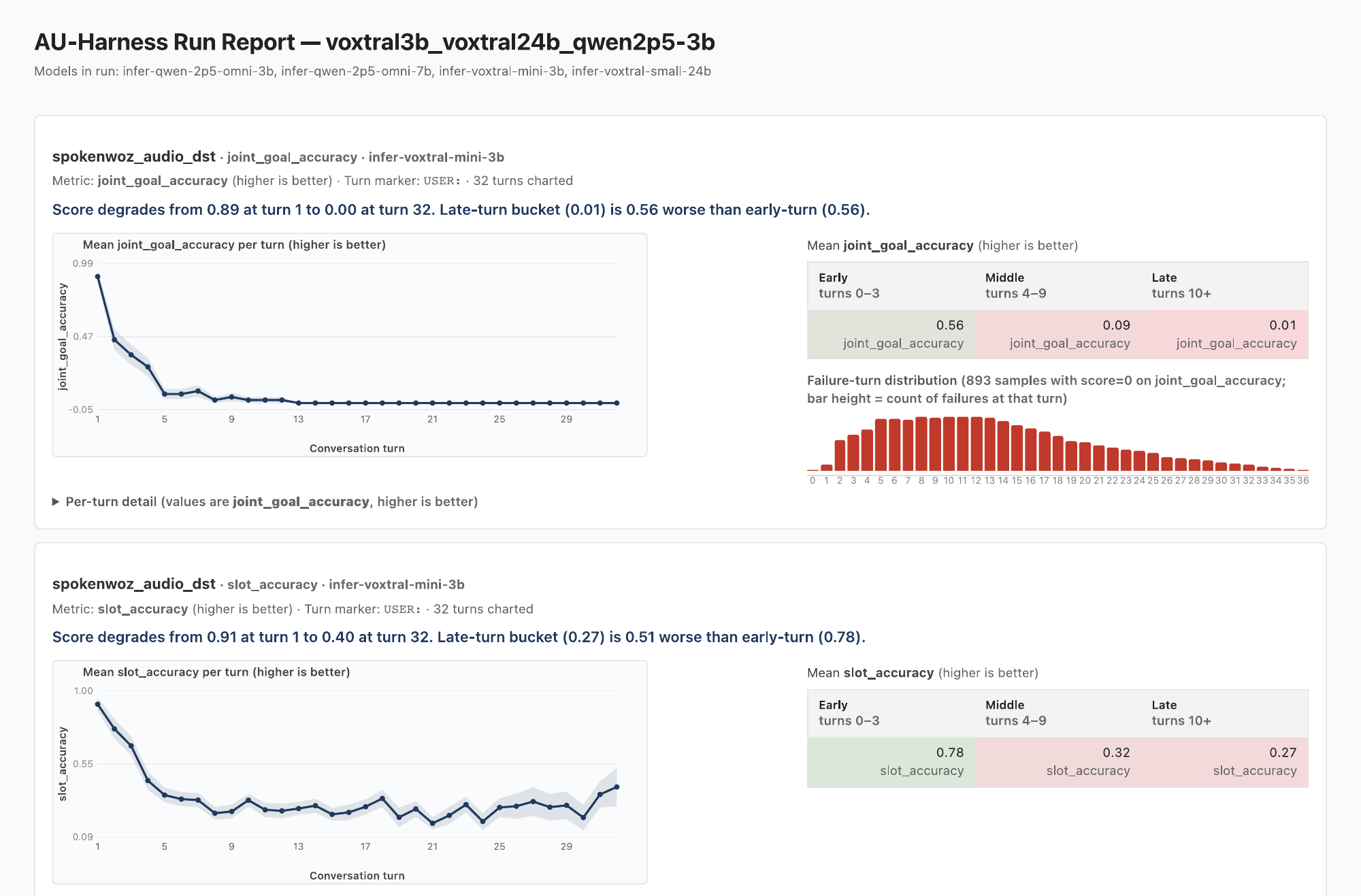}
    \caption{\textbf{Multi-turn.} Multi-turn targets long-conversation degradation, currently instrumented on SpokenWOZ-style dialogue state tracking with joint goal accuracy, slot accuracy, and slot F1. For each (task, metric, model) triple the tab plots mean score per turn, summarizes early (turns 0–3), middle (4–9), and late (10+) buckets, and renders a failure-turn histogram showing which turn indices contribute the most zero-score samples. Bucket cells are tinted by absolute level, making floor effects in the late bucket visually distinguishable from a smooth glide.}
    \label{fig:rr-multi-turn}
\end{figure}

\begin{figure}[t]
    \centering
    \includegraphics[width=1\linewidth]{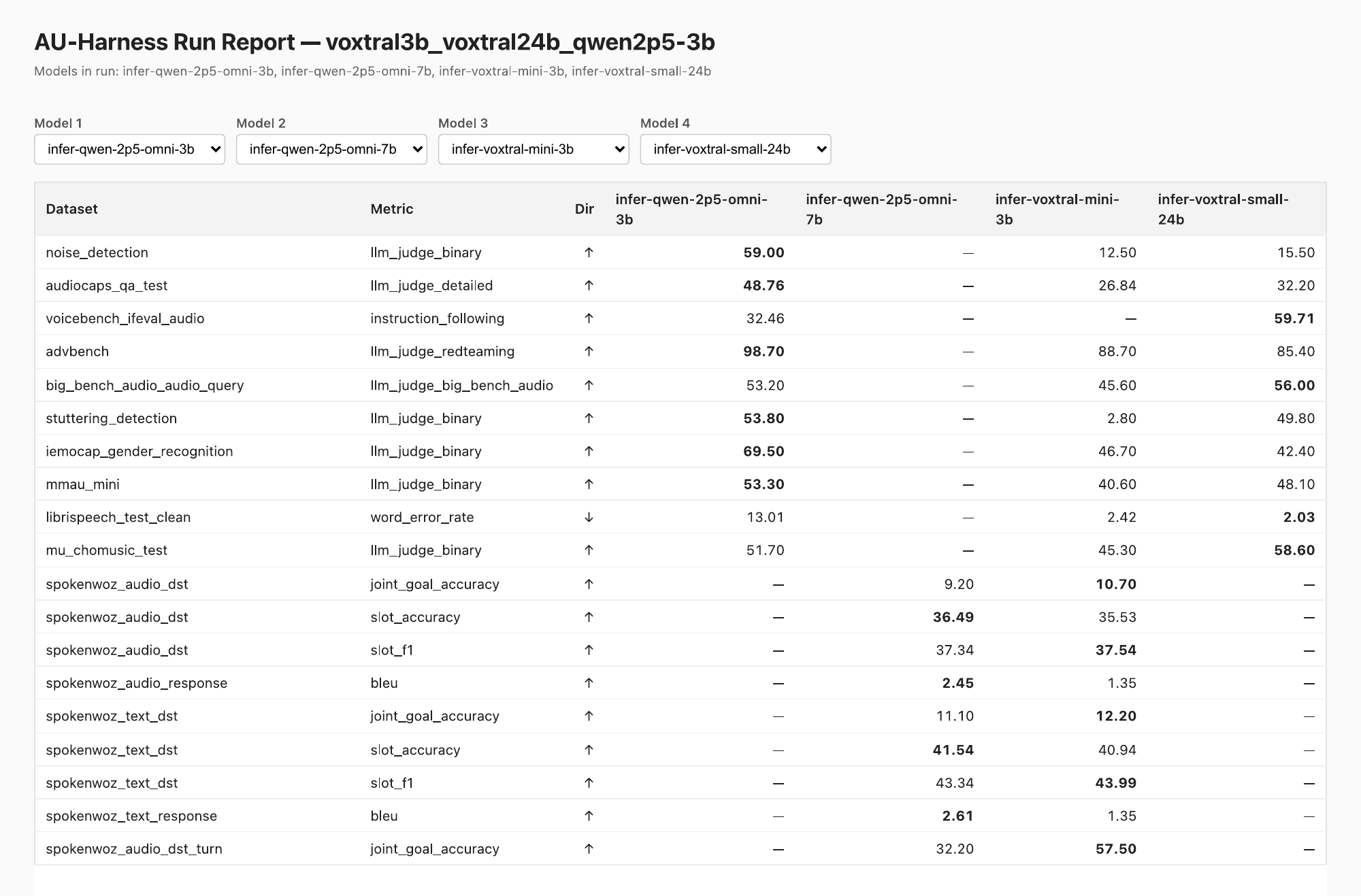}
    \caption{\textbf{Compare.} Compare is a head-to-head view that accepts up to four model selections and renders per-task scores side by side. A Dir column encodes metric direction (↑ higher is better; ↓ lower is better, e.g. word error rate), and the leading score per row is bolded. The header line summarizes wins by aggregating direction-aware comparisons across the tasks where every selected model has a valid score; tasks with any missing score are excluded from the win count but remain visible in the table.}
    \label{fig:rr-compare-models}
\end{figure}

\begin{figure}[t]
    \centering
    \includegraphics[width=1\linewidth]{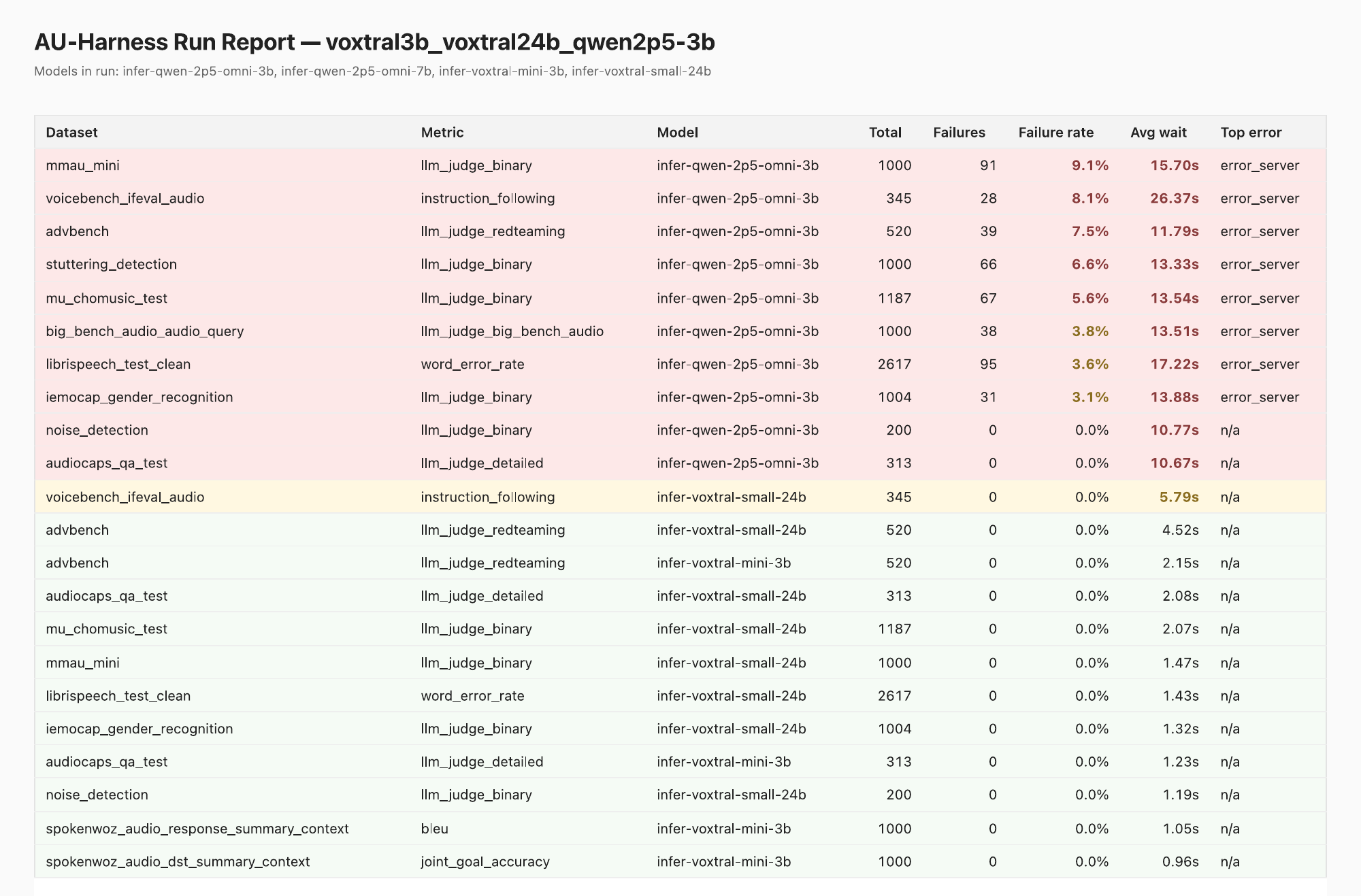}
    \caption{\textbf{Ops Health.} Ops health separates inference reliability from model quality. For each (dataset, metric, model) row it reports the total request count, hard-failure count, failure rate, mean wall-clock wait per request, and the most common error class. Highlighted rows surface transient API outages, rate-limit storms, or per-model timeout misconfigurations that would otherwise be mis-read as quality regressions in the other tabs.}
    \label{fig:rr-ops-health}
\end{figure}

The AU-Harness run report is a self-contained HTML artifact constructed at the end of every evaluation run. It surfaces the same underlying records used to populate the leaderboard, but reorganizes them around the questions practitioners typically ask once a run completes: which categories went well, where did regressions land, what failed at the infrastructure layer, and long-context behavior analysis for multi-turn benchmarks. The report exposes seven tabs, summarized below; Figures \ref{fig:rr-overview}, \ref{fig:rr-config}, \ref{fig:rr-categories}, \ref{fig:rr-patterns}, \ref{fig:rr-multi-turn}, \ref{fig:rr-compare-models}, \ref{fig:rr-ops-health} show each one for a run config with Qwen2.5-Omni-3B and 7B, Voxtral-Mini-3b, and Voxtral-Small-24B.

\subsection{Turn Range Selection} \label{sec:turn-range-selection} We cap evaluation at 20 turns because the all-audio configurations (where each user turn is raw audio alongside agent turn as text) approach the context length limits of current audio language models, and can lead to truncation beyond this point. While text-only configurations could in principle extend further, we fix the window across all settings to ensure comparable and consistent evaluation. Additionally, we omit the first turn from all plots as the opening turn is nearly always a greeting that introduces no new slots and adds visual noise without analytical value, especially for SpokenWOZ.

\subsection{Licensing} 
\label{app:license}
\framework~is released under Apache-2.0 license. The framework does not host or contain datasets; therefore, running evaluations of certain datasets with \framework~follow the licenses of the original datasets' publishers and the aforementioned \framework' license. 

\subsection{User Case Study}
 To assess the ease-of-adoption of our \framework, we conduct a controlled user case study. More specifically, our study involves 5 participants experienced in LLM evaluations and the users are expected to leverage AU-Harness and exend it to new tasks and model from scratch. Following~\citet{joshi2025coprompter,ismanto2026improvement}, we compute System Usability Scale (SUS) and Technology Acceptance Model (TAM) assessment scores.

 \paragraph{Participant Background} All selected participants had prior hands-on experience with at least one of the contemporary LLM evaluation toolkits. This decision is deliberate to ensure the fair comparative assessment of the \framework' usability.

 \paragraph{Guideline for participants} We evaluate \framework~through the following realistic workflows when utilizing AU-Harness and extending it to new task(s) and new model(s).

 \begin{enumerate}[nolistsep, noitemsep, topsep=0pt, partopsep=0pt, leftmargin=*]
 \item Install and configure the toolkit directly from source code
\item Set up the evaluation environment by reviewing README documentation (specifically \textit{run\_config}, \textit{task\_config} definitions and specifications)
\item Create \textit{task\_config} to integrate the specified task into the toolkit
\item Introduce the model configurations to the \textit{run\_config}
\item Execute the benchmark with the toolkit
\item Inspect result logs and validate successful completion
\end{enumerate}

To better the scope the study, the target model and task are specified upfront, enabling the focus on integrating them into \framework. For task, we select MMAU-PRO (music subset)~\cite{kumar2026mmau} and for the model, we leverage Qwen3-Omni-30B-A3B-instruct~\cite{xu2025qwen3}.

\paragraph{Observations} The observed results are strongly positive among participants of \framework~user study. SUS score is averaged 87.5/100 (an excellent rating above the industry benchmark) indicating participants found \framework~genuinely easy to install, configure, and operate despite the complexity of the workflow. TAM score is averaged 4.67/5, reflecting strong perceived usefulness, ease of use, and a clear intention to adopt and recommend the tool for future LALM evaluations. Together, these scores suggest \framework~delivers a smooth setup experience and real research value, with participants encountering few usability barriers when extending it to new models and benchmarks. Because the study tests the tool rather than the participants, any friction reported is treated as actionable feedback for improving the tool design.

%% file: tables/08_benchmark_details_v2.tex
\begin{table*}[t]
\resizebox{\textwidth}{!}{%
\begin{tabular}{@{}cllcll@{}}
\toprule
\multicolumn{1}{c}{\textbf{Task   Category}} & \textbf{Task Type} & \textbf{Dataset Name} & \textbf{Task} & \textbf{Description} & \textbf{License} \\ \midrule
\multirow{18}{*}{Speech Recognition} & ASR & AISHELL-1 & 1 & High-quality Mandarin speech recognition dataset & Apache 2.0 \\
 & ASR & AMI Meeting Corpus & 2 & Multispeaker meeting recordings for ASR and diarization & CC BY 4.0 \\
 & ASR & CallHome & 5 & Conversational speech corpus across multiple languages & LDC User Agreement for Non-Members \\
 & ASR & Common Voice & 100 & Crowdsourced multilingual speech dataset from Mozilla & CC0 1.0 Universal \\
 & ASR & FLEURS EN-US & 102 & Multilingual speech dataset for ASR and translation & CC BY 4.0 \\
 & ASR & GigaSpeech & 1 & Large-scale audio and transcription corpus for end-to-end ASR & Apache 2.0 \\
 & ASR & GigaSpeech2 & 2 & Large-scale audio and transcription corpus for end-to-end ASR (v2) & Apache 2.0 \\
 & ASR & LibriSpeech & 2 & Audiobook-derived speech corpus with clean and noisy subsets & CC BY 4.0 \\
 & ASR & Multilingual LibriSpeech (MLS) & 7 & Extension of LibriSpeech with multiple European languages & CC BY 4.0 \\
 & ASR & MNSC & 6 & Large-scale multitask speech corpus & MNSC: Publicly released \\
 & ASR & People's Speech & 1 & Large-scale open-source English speech recognition dataset & CC-BY-SA \\
 & ASR & SPGISpeech & 1 & Transcriptions of financial meeting recordings & Kensho User Agreement \\
 & ASR & TEDLIUM3 & 1 & Transcribed TED talks for ASR and speaker adaptation & CC BY-NC-ND 3.0 \\
 & ASR & VoxPopuli & 17 & Multilingual speech corpus from European Parliament recordings & CC0 \\
 & Code-switching ASR & SEAME & 2 & Mandarin-English code-switching speech dataset & LDC2015S04 \\
 & Long-form ASR & TEDLIUM3 & 1 & Transcribed TED talks for ASR and speaker adaptation (long-form version) & CC BY-NC-ND 3.0 \\
 & Long-form ASR & Earnings21 & 1 & Long-form earnings call dataset for speech recognition & CC-BY-SA-4.0 \\
 & Long-form ASR & Earnings22 & 1 & Long-form earnings call dataset for speech recognition & CC-BY-SA-4.0 \\
 \midrule
\multirow{11}{*}{Paralinguistics} & Accent Recognition & MNSC AR Dialogue & 1 & Dataset for accent recognition in dialogue speech & MNSC: Publicly released \\
 & Accent Recognition & MNSC AR Sentence & 1 & Dataset for accent recognition in sentence-level speech & MNSC: Publicly released \\
 & Accent Recognition & VoxCeleb Accent & 1 & Speech dataset with diverse speakers for accent recognition & CC BY 4.0 \\
 & Emotion Recognition & IEMOCAP Emotion & 1 & Multimodal dataset for emotion recognition in speech & GPL-3.0 \\
 & Emotion Recognition & MELD Emotion & 1 & Multi-party conversation dataset for emotion recognition & GPL-3.0 \\
 & Emotion Recognition & MELD Sentiment & 1 & Multi-party conversation dataset for sentiment analysis & GPL-3.0 \\
 & Gender Recognition & IEMOCAP Gender & 1 & Multimodal dataset for gender recognition in speech & GPL-3.0 \\
 & Gender Recognition & MNSC GR Dialogue & 1 & Dataset for gender recognition in dialogue speech & MNSC: Publicly released \\
 & Gender Recognition & MNSC GR Sentence & 1 & Dataset for gender recognition in sentence-level speech & MNSC: Publicly released \\
 & Gender Recognition & VoxCeleb Gender & 1 & Speech dataset with diverse speakers for gender recognition & CC BY 4.0 \\
 & Speaker Recognition & MMAU-mini & 1 & Multi-modal audio dataset for speaker recognition & Apache 2.0 \\
 \midrule
\multirow{6}{*}{Audio Understanding} & Music Understanding & MuChoMusic & 1 & Benchmark for music understanding for LALMs & CC-BY-SA-4.0 \\
 & Scene Understanding & AudioCaps & 1 & Large-scale dataset for open-domain audio captioning & MIT \\
 & Scene Understanding & AudioCaps QA & 1 & Dataset for question answering over natural audio scenes & MIT \\
 & Scene Understanding & Clotho AQA & 1 & Dataset for answering natural-language questions about audio signals & MIT \\
 & Scene Understanding & WavCaps & 1 & Large-scale weakly labeled dataset for audio captioning & CC-BY-NC 4.0 \\
 & Scene Understanding & WavCaps QA & 1 & Large-scale dataset for audio question answering & CC-BY-NC 4.0 \\
 \midrule
\multirow{14}{*}{Spoken Language Understanding} & Intent Classification & SLURP & 1 & Multi-domain spoken dialogue understanding benchmark & CC BY-NC 4.0 \\
 & Speech QA & Alpaca Audio & 1 & Speech dataset for question answering with audio instructions & Apache-2.0 \\
 & Speech QA & CN College Listen MCQ & 1 & Multispeaker dataset for listening-based multiple-choice questions & MERaLiON Public License \\
 & Speech QA & Dream TTS MCQ & 1 & Dialogue-based multiple-choice comprehension dataset with audio & MIT \\
 & Speech QA & MNSC SQA & 4 & Benchmark for reasoning and understanding in spoken language & NSC License \\
 & Speech QA & OpenHermes & 1 & Speech dataset for question answering with audio instructions & CC-BY-NC \\
 & Speech QA & Public-SG & 1 & Speech question answering benchmark & NSC License \\
 & Speech QA & SLUE SQA & 1 & Spoken Language Understanding Evaluation benchmark & CC-BY-4.0 \\
 & Speech QA & Spoken Squad & 1 & Speech dataset for extraction-based question answering & CC-BY-SA-4.0 \\
 & SQQA & Big Bench Audio & 2 & Benchmark for reasoning with audio and text input & MIT \\
 & SQQA & MMSU & 12 & Multi-choice question answering dataset & Apache-2.0 \\
 & SQQA & OpenBookQA & 1 & Multi-choice question answering dataset & Apache-2.0 \\
 & SQQA & SD-QA & 22 & Multi-choice question answering dataset & Apache-2.0 \\
 & Translation & CoVoST2 (zh$\rightarrow$en) & 36 & Large-scale multilingual dataset for speech translation & CC-BY-NC-4.0 \\
 \midrule
 \multirow{6}{*}{Spoken Language Reasoning} 
 & Speech Instruction Following & IFEVAL & 2 & Speech and text-equivalent dataset for complex instruction following & Apache-2.0 \\
 & Speech Instruction Following & VoiceBench-MTBench (46/80 samples) & 1 & Speech dataset for multi-turn instruction following & Apache-2.0 \\
 
 & Grade School Math & GSM8k & 2 & Speech and text-equivalent dataset of grade school math word problems & MIT (text dataset) \\





& Speech Function Calling & BFCL\footnotemark[1] & 20 & Speech and text-equivalent dataset for complex function calling tasks & Apache-2.0 \\
& Speech Instruction Following & MTBench\footnotemark[2] & 2 & Speech and text-equivalent dataset for multi-turn instruction following & Apache-2.0 \\
&Speech-to-Coding & SPEECH\_TO\_SQL\footnotemark[3] & 2 & Speech and text-equivalent dataset for generating executable SQL code & Apache-2.0 \\
 \midrule


\multirow{2}{*}{Safety \& Security} & Safety & Advbench & 1 & Speech dataset for testing resistance to adversarial or harmful prompts & Apache 2.0 \\
 & Spoofing & ASVpoof2017 & 1 & Speech dataset for spoofing attack detection in real-world conditions & CC BY-NC 4.0 \\

 \midrule
\multicolumn{1}{l}{} &  & \textbf{Total Tasks} & \textbf{387} &  &  \\ \bottomrule
\end{tabular}%
}
\centering
\caption{\textbf{Comprehensive Audio and Speech Datasets Overview.} Listing of 61 datasets across 6 task categories: Speech Recognition, Paralinguistics, Audio Understanding, Spoken Language Understanding, Spoken Language Reasoning, and Safety \& Security.}
\label{tab:audio_speech_datasets}
\end{table*}

\footnotetext[1]{\url{https://huggingface.co/datasets/ServiceNow-AI/BFCL\_v3\_audio}}
\footnotetext[2]{\url{https://huggingface.co/datasets/ServiceNow-AI/mt-bench\_audio}}
\footnotetext[3]{\url{https://huggingface.co/datasets/ServiceNow-AI/spider\_audio}}

%% file: tables/08_ablation_efficiency_eval_settings.tex
\begin{table*}[htb]
\small
\resizebox{0.8\textwidth}{!}{%
\begin{tabular}{ccccc}
\toprule
 & \textbf{MELD-Emotion} & \textbf{Librispeech-clean} & \textbf{ClothoAQA} \\
 \midrule
\textbf{\# Samples} & 500 & 500 & 500 \\
\textbf{Audio Duration (seconds)} & 1,476 & 3,780 & 11,376 \\
\textbf{Evaluation Metric} & LLM-Judge & WER & LLM-Judge \\
\bottomrule
\end{tabular}%
}
\captionsetup{font=small}
\centering
\caption{\textbf{Experimental setup for efficiency comparison across evaluation frameworks.} We conduct controlled experiments using 500 samples from three diverse datasets: MELD-Emotion (short emotional speech), LibriSpeech-clean (medium-length read speech), and ClothoAQA (long-form descriptive audio). Total audio duration varies from 1,476 to 11,376 seconds, enabling assessment across different audio characteristics and evaluation modalities (LLM-judge vs. traditional metrics).}
\label{tab:ablation_efficiency_eval_settings}
\end{table*}

%% file: layout_figures/efficiency_ablations.tex
\begin{figure*}[htb]
    \captionsetup{font=small}
    \centering
    \begin{subfigure}{0.32\textwidth}
        \centering
        \includegraphics[width=\linewidth]{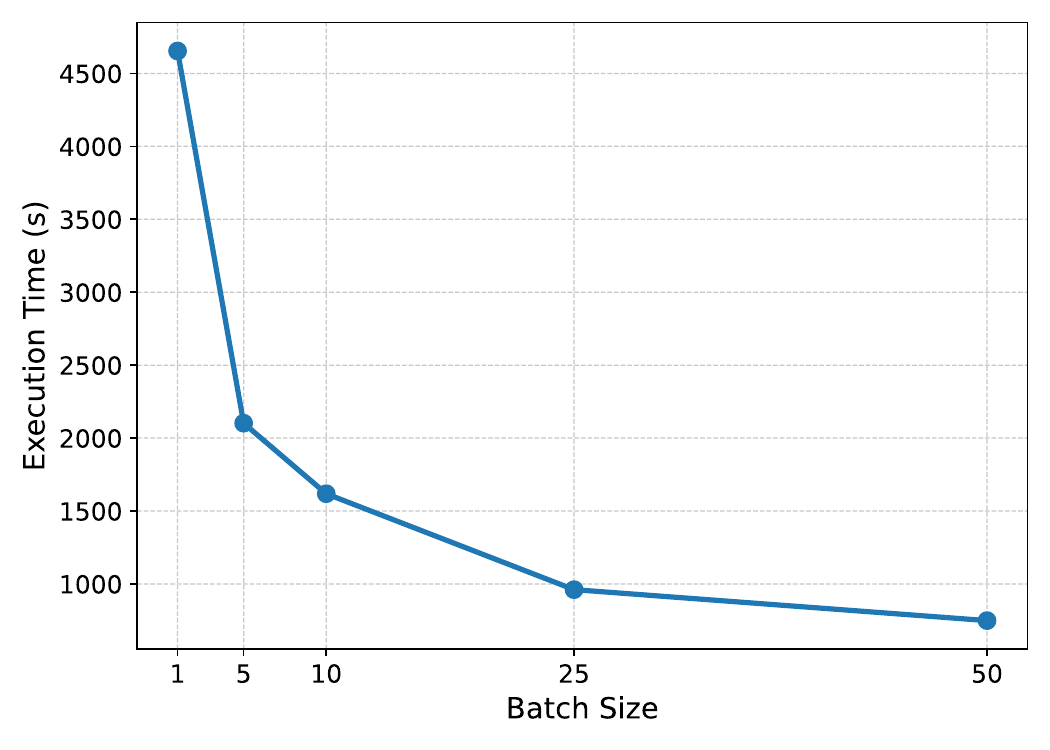}
        \caption{Batch size effect}
        \label{fig:batchsize}
    \end{subfigure}
    \hfill
    \begin{subfigure}{0.32\textwidth}
        \centering
        \includegraphics[width=\linewidth]{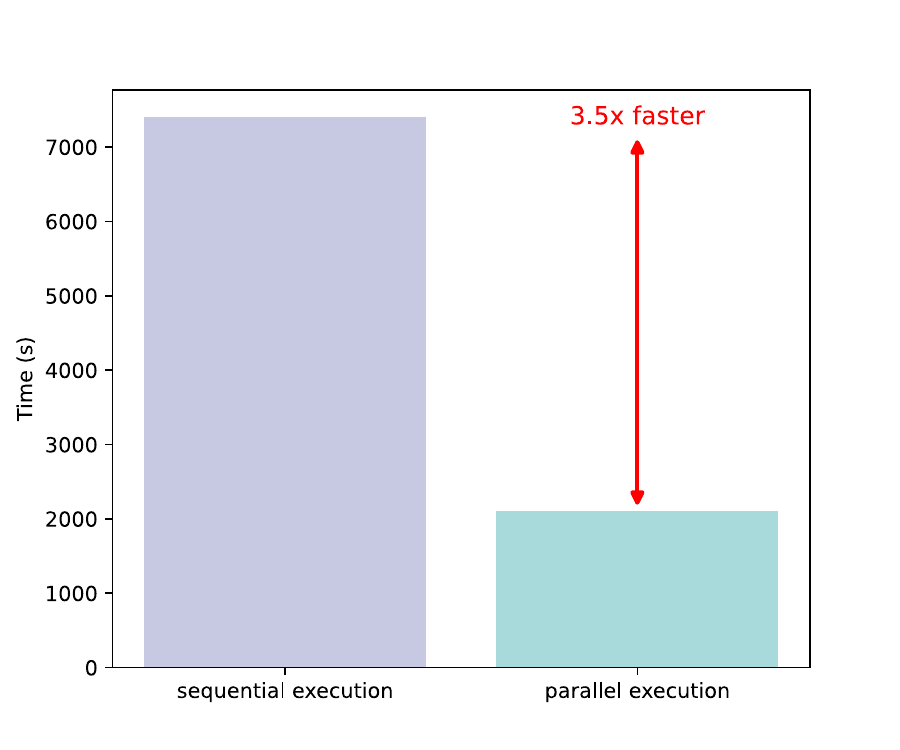}
        \caption{Throughput scaling}
        \label{fig:throughput}
    \end{subfigure}
    \hfill
    \begin{subfigure}{0.32\textwidth}
        \centering
        \includegraphics[width=\linewidth]{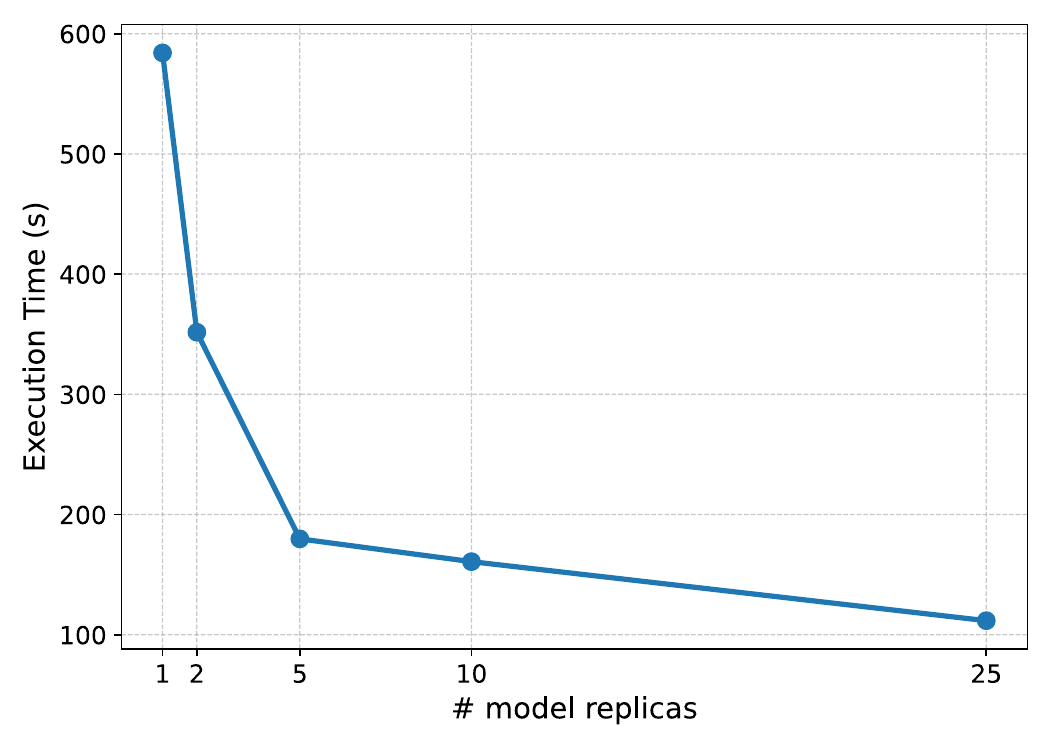}
        \caption{Latency trade-offs}
        \label{fig:latency}
    \end{subfigure}
    \caption{\textbf{ Inference efficiency ablations in \framework{}.} 
    We examine three factors: (a) impact of batch size on execution time, (b) throughput gains from parallel execution, and (c) latency reduction through replica scaling.}
    \label{fig:3_efficiency_experiments}
\end{figure*}